\documentclass[12pt]{iopart} 

\usepackage{graphicx}
\usepackage{amssymb}
\usepackage{amsfonts}
\usepackage{bm}
\usepackage{cite}

\begin{document}
\jl{2}
\renewcommand {\d} {{\rm d}}
\renewcommand {\i} {{\rm i}}

\newcommand {\balpha} {\boldsymbol{\alpha}}
\newcommand {\bbeta}  {\boldsymbol{ \beta}}
\newcommand {\bgamma} {\boldsymbol{ \gamma}}
\newcommand {\bnabla} {\boldsymbol{ \nabla}}
\newcommand {\bsigma} {\boldsymbol{ \sigma}}

\newcommand {\ee} {{\rm e}}

\newcommand {\bfa} {{\bf a}}
\newcommand {\bfb} {{\bf b}}
\newcommand {\bfe} {{\bf e}}
\newcommand {\bfk} {{\bf k}}
\newcommand {\bfn} {{\bf n}}
\newcommand {\bfp} {{\bf p}}
\newcommand {\bfq} {{\bf q}}
\newcommand {\bfr} {{\bf r}}
\newcommand {\bfv} {{\bf v}}

\newcommand {\bfA} {{\bf A}}
\newcommand {\bfP} {{\bf P}}
\newcommand {\bfQ} {{\bf Q}}
\newcommand {\bfR} {{\bf R}}
\newcommand {\bfY} {{\bf Y}}

\newcommand {\cF} {{\cal F}}
\newcommand {\cO} {{\cal O}}
\newcommand {\cP} {{\cal P}}
\newcommand {\cT} {{\cal T}}
\newcommand {\cZ} {{\cal Z}}

\newcommand {\rp} {{\rm p}}
\newcommand {\Q} {Z_{\rp} e}

\newcommand {\E} {{\varepsilon}}
\newcommand {\om} {{\omega}}
\newcommand {\Om} {{\Omega}}
\newcommand {\tom} {{\tilde{\omega}}}
\newcommand {\tth} {{\tilde{\theta}}}

\newcommand {\qmin} {q_{\min}}
\newcommand {\qmax} {q_{\max}}

\newcommand {\tot} {{\rm tot}}
\newcommand {\pol} {{\rm pol}}
\newcommand {\ord} {{\rm ord}}
\newcommand {\rint} {{\rm int}}
\newcommand {\NR} {{\rm NR}}


\newcommand {\hc} {\dagger}

\newcommand {\threejot}[6]{\pmatrix{ #1\!\!&#2\!\!&#3\!\cr #4&#5&#6  \cr}}

\title{Relativistic effects in polarizational bremsstrahlung.}


\author{A.~V.~Korol$^{\rm a}$, A.~V.~Solov'yov$^{\rm b}$
\footnote{On leave from: Ioffe Physical-Technical Institute, 
Russian Academy of Sciences, St. Petersburg 194021, Russia}
}

\address{
     $^{\rm a}$\, 
     Department of Physics, Russian Maritime Technical University, 
     Leninskii prospect 101, St. Petersburg  198262, Russia} 
  
\address{$^{\rm b}$\, 
     Frankfurt Institut for Advanced Studies,
     Johann Wolfgang Goethe-Universit\"at, 
     60054 Frankfurt am Main, Germany}

\begin{abstract}
We review the achievements of the theory of polarization bremsstrahlung 
of relativistic particles, including the case when both colliders have 
internal structure.
The main features which the relativistic effects bring into the
problem are discussed and illustrated by the results of 
numerical calculations.
\end{abstract}

\section{Introduction.}
\label{Introduction}

In this paper we review the results of theoretical studies of the 
polarizational bremsstrahlung (PBrS) of relativistic
atomic particles.
Another review article from this issue \cite{NonRelBrS_2004}
is devoted to the discussion of the PBrS process during various 
non-relativistic collisions.
There a comprehensive historical review is given, and
most of the terminology which is used to describe the PBrS process 
is introduced.
Therefore, wherever possible we refer to \cite{NonRelBrS_2004}
in order to avoid the repetition.

In what follows we focus on the specific features of PBrS 
(see figure 1b in \cite{NonRelBrS_2004})
which are due to the relativistic effects.
The latter can be subdivided into the following categories.
Firstly, there are effects directly related to the 
relativistic velocities of the particles involved in the process:
{\em the velocity of a projectile} and/or the 
{\em  the velocity of the orbital motion} of electrons inside the colliders. 
To account for these effects one has to describe the dynamics
of the colliders using the Dirac equation rather than
the Schr\"odinger equation.

The relativistic effect of {\em retardation}, which modifies 
the interaction between a projectile and a target,
 we attribute to another category.
Retardation implies that the relativistic particle polarizes the target
not only through the Coulomb field, but also via the field of the 
transverse virtual photons  (e.g. \cite{Akhiezer}), or, in terms of 
classical electrodynamics, via the retarded vector potential 
\cite{Landau2}.
The effective radius of this field increases infinitely as
the velocity $v_1$ of the projectile approaches the velocity of light $c$.
Therefore, the ultra-relativistic charge, $v_1\approx c$,
polarizes the target
mainly via the exchange of the transverse virtual photons.

There are relativistic effects due to the {\em multipole character} 
of radiation emitted by the target electrons.
The multipolarity of the PBrS radiation depends solely on the
magnitude of $k R_{\rm at}$, where $k$ is the photon momentum and
$R_{\rm at}$ is the (mean) radius of the orbit of the electron who 
radiates.
If this parameter is small, then the dipole-photon approximation
is applicable, if otherwise, then it is necessary to take into
account the radiation in higher multipoles.

Finally, in the case when both colliders are complex particles
(for example, atom-atom, atom-ion, ion-ion pairs) 
it is necessary to account for the relativistic Doppler effect and
the abberation of radiation.
These are important factors which modify the PBrS of a relativistic
complex projectile.

The role of the relativistic effects mentioned above was studied
in
\cite{AmusiaKorolKuchievSolovyov1985,
AstapenkoBuimistrovKrotovMikhailovTrahtenberg1985,
AmusiaKuchievSolovyov1987,AmusiaSolovyov1988,AmusiaSolovyov1990a,
AvdoninaPratt1999,AstapenkoBureevaLisitsa2000b,
KorolLyalinObolenskySolovyovSolovjev2001,
OurRelativisticJPB,OurRelativisticJETP,
KorolObolenskySolovyovSolovjev2002}
in various collisions of isolated atomic particles.
In this paper we review the results obtained in the cited papers. 

The influence of a dense medium on the features 
of PBrS of relativistic electrons was investigated in 
\cite{BlazhevichEtal1996,Nasonov1998,BlazhevichEtal1999,
KamyshanchenkoNasonovPokhil2001,AstapenkoBuimistrovKrotovNasonov2004}.
This issue is beyond the scope of our review and is not discussed below.
We do not discuss the properties of the ordinary BrS
of relativistic particles.
The list of corresponding references (although not the fullest one) 
can be found \cite{NonRelBrS_2004,OurRelativisticJPB}.
Finally, the main topic of this review concerns the relativistic
effects in the PBrS process, therefore, we omitted the description
of general features of PBrS  in the non-relativistic domain
as well as the historical survey.
These could be found in \cite{NonRelBrS_2004,AstapenkoBureevaLisitsa2002}.

In the first two papers on relativistic PBrS formed in a collision of
a charged structureless particle with an isolated atom (ion)
\cite{AmusiaKorolKuchievSolovyov1985,
AstapenkoBuimistrovKrotovMikhailovTrahtenberg1985}
the process was treated within the framework of 
the relativistic plane-wave Born approximation for the projectile and 
in the non-relativistic approximation for the target electrons.
The retardation effect was accounted for, whereas the emission of the
photon via the polarizational mechanism was considered in the dipole-photon
approximation.
The following two important features were established in these papers.
Firstly, it was demonstrated that the PBrS cross section increases
logarithmically with the energy $\E_1$ of a projectile.
This feature is a result of the retardation in the projectile--atom
interaction.
Secondly, it was shown that the shape of the angular distribution of PBrS,
in contrast to the ordinary BrS, is proportional to 
$(1+a\cos^2\theta)$ ($\theta$ is the emission angle) 
where the coefficient $a$ weakly depends on $\E_1$.
This feature reflects the nature of the PBrS process in which
the radiation occurs due to the alteration of the 
induced dipole moment of the target during the collision.
We discuss these features in more detail in section  \ref{Structureless}.

Also, it was demonstrated \cite{AmusiaKorolKuchievSolovyov1985} 
that the amplitude $f_{\pol}$ of the dipole PBrS 
can be expressed in terms of  two polarizabilities, 
which define the dynamic atomic response to the 
joint actions of the fields of the projectile and of the dipole photon.
In addition to the polarizability $\alpha(\omega,q)$, 
which accounts for the (non-retarded) Coulomb  interaction 
between the projectile and the target, and which defines the 
PBrS of a non-relativistic projectile (see Eq. (1) in \cite{NonRelBrS_2004}),
there is another polarizability, $\beta(\omega,q)$,
which describes the dynamic response to the retarded vector potential
created by the projectile.
In the limit $q \ll R_{\rm at}^{-1}$ both polarizabilities
are expressed via the dynamic dipole polarizability $\alpha_{\rm d}(\omega)$
of the target.
Small values of the transferred momenta $q$ correspond to large distances
between the projectile and the target, $r\gg R_{\rm at}$,
which are important for the PBrS mechanism 
\cite{Zon1977,AmusiaZimkinaKuchiev1982}.
This fact, known from the non-relativistic theory of PBrS, was utilized
in  \cite{AmusiaKorolKuchievSolovyov1985,
AstapenkoBuimistrovKrotovMikhailovTrahtenberg1985}
to derive the formulae for the PBrS spectrum and angular 
distribution within the large-distance approximation
(another term is `the logarithmic approximation'
\cite{Zon1977,AmusiaZimkinaKuchiev1982}).
In \cite{AstapenkoBureevaLisitsa2000b} the `logarithmic'
approximation was applied to study the total BrS process 
(i.e. the polarizational and the ordinary channels)
in combination with the use of the local density 
method for the computation of $\alpha_{\rm d}(\omega)$.

In \cite{AvdoninaPratt1999} an approach was suggested for
calculating the total BrS spectrum in relativistic electron-atom scattering
utilizing a relativistic  modified Elwert-Born approximation for the 
ordinary component of the spectrum but considering the PBrS within
the  non-relativistic 'stripping' approximation 
\cite{BuimistrovTrakhtenberg1977,AmusiaAvdoninaChernyshevaKuchiev1985}.

The theory of PBrS formed in a collision of a charged particle with 
a many-electron atom/ion was developed further in the papers 
\cite{KorolLyalinObolenskySolovyovSolovjev2001,
OurRelativisticJPB,OurRelativisticJETP,
KorolObolenskySolovyovSolovjev2002} where the fully relativistic
formalism and the results of numerical calculations were
presented.
The formalism developed in these  papers accounts for the relativistic
effects of all types mentioned above (with the exception for the Doppler
shift and abberation which are irrelevant in the case of a structureless
projectile).
The approach was based on the use of the distorted partial wave formalism
for the description of the scattering process.
The states of the target's electrons were described in terms of 
single-electron relativistic wavefunctions.
The effect of 
retardation and the multipole expansion of the emitted photon wave
were accounted for.
It was demonstrated that the full relativistic PBrS amplitude
is expressed in terms of the multipole generalized target
polarizabilities of three different types corresponding to the allowed
combinations of the types of virtual and  emitted photons 
coupled in the amplitude.
On the basis of the relativistic formulae for the amplitude and 
the cross section obtained in \cite{OurRelativisticJPB,OurRelativisticJETP}
the results, which had been obtained within the frameworks of 
the various simpler approaches, were evaluated 
by considering the corresponding limiting procedures.
It was shown that the relativistic effects lead to noticeable
changes in the spectral and angular characteristics of PBrS.
A more detailed discussion is presented in section  \ref{Structureless}. 

The PBrS mechanism defines the emission spectrum in collision
of two complex particles, such as atoms or ions.
The relativistic theory of such processes was developed  in 
\cite{AmusiaKuchievSolovyov1987,AmusiaSolovyov1988,AmusiaSolovyov1990a}
where the details of the formalism were presented and the analysis
of a number of physical phenomena and limiting cases was carried out.
The important difference between the radiation formed in the atom--atom
(or ion--ion/atom) collision from the collision of a structureless 
particle with an atom/ion is that in the former case both colliders
emit photons via the polarization mechanism (the ordinary BrS of the 
projectile is negligibly small because of the large mass).
The PBrS of the target, which is at rest in a laboratory frame, 
bears the same features as in the particle-atom/ion collision.
On the contrary, the spectrum and the angular distribution of 
PBrS emitted by a relativistic complex projectile is strongly 
influenced by  the relativistic Doppler shift and the abberation 
of radiation. 
This influence manifests itself differently, depending on the magnitude of the
photon energy $\om$ measured in the rest frame of the projectile.
For high values of  $\om$ the radiation, measured in the laboratory frame,
is concentrated  in the cone $\theta \sim \gamma^{-1}$ 
($\gamma$ is the relativistic Lorenz factor of the projectile).
For small values of $\om$ the angular distribution is nearly isotropic.
Therefore, as it was pointed out in
\cite{AmusiaSolovyov1988,AmusiaSolovyov1990a}
for sufficiently high velocities of the collision there is
a principal possibility to separate the radiation from the two colliders.
Another important feature of the BrS process at relativistic velocities,
which is also due to the Doppler and the abberation effects, is
that the intensity of the dipole radiation formed in symmetric collisions
does not vanish.
In some more detail these and some other effects are discussed in 
section \ref{AtomAtom}.

\section{Polarizational BrS in collisions of structureless
particles with atoms.}
\label{Structureless}

\begin{figure}
\begin{center}
\includegraphics[scale=0.8]{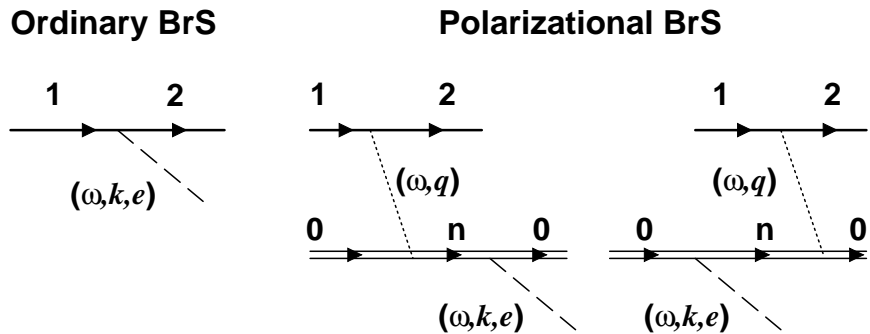}
\end{center}
\caption{Diagrammatical representation of the BrS process (ordinary and
polarizational) for a relativistic structureless charged projectile 
scattered by a many-electron atom.
The solid lines correspond to the relativistic projectile which
moves in a central field of the target.
The initial ('1') and the final ('2')  states of the 
projectile are characterized by the asymptotic momenta $\bfp_{1,2}$ and 
the polarizations $\mu_{1,2}$.
The double lines denote the states of the target: index '0' marks
the initial and the final states, index '$n$' corresponds to the
intermediate virtual state.
The dashed lines designate the emitted (real) photon of energy
$\om$, momentum $\bfk$ and the polarizational vector $\bfe$.
The dotted lines stand for the virtual photon, the energy of
which is also $\om$ but the momentum $\bfq$ is not fixed by any
kinematic relations.}
\label{diagram1.fig}
\end{figure}

The BrS process in the collision between a relativistic 
projectile of charge $Z_{\rp} e$ and mass $m$ and a target atom 
is the transition of the projectile from the initial state
$(\bfp_1,\mu_1)$ to the final state $(\bfp_2,\mu_2)$ 
accompanied by the emission of the photon.
The energies of the particle in the initial and the final states
are found from $\E_{1,2} = (p_{1,2}^2 + m^2 )^{1/2}$
(in this section we use the relativistic system of units $\hbar=m_e=c=1$).
This radiative transition can occur via two mechanisms, as
it is illustrated by 
the Feynman diagrams presented in figure \ref{diagram1.fig}.
The first diagram describes  the ordinary BrS amplitude, $f_{\ord}$.
The two other diagrams represent the polarizational part of the amplitude,
$f_{\pol}$.

The relativistic formalism of the derivation of the OBrS
amplitude within the first Born approximation, or a more 
sophisticated one, based on the use of Sommerfeld-Maue functions, 
can be found in textbooks (e.g.  \cite{Land4,Akhiezer}).
The description in the framework of the distorted partial-wave 
approximation (DPWA) is also available \cite{PrattTseng1970,Tseng1997}.

In what follows we outline the principal steps of the evaluation of
the PBrS amplitude $f_{\pol}$ focusing on the differences between
the full relativistic description from the non-relativistic one.
For the sake of clarity we assume that the spin of a projectile 
is equal to $1/2$, and, therefore, its bi-spinor wavefunction 
satisfies the Dirac equation which accounts for the central field of 
the target. 
It is a not severe restriction, and all the formulae can be easily 
re-written for the relativistic projectiles of other value of the spin.

As in the non-relativistic case, the photon emission via the polarizational 
mechanism occurs due to the virtual excitations of the target electrons 
under the action of two fields: the field created by the charged projectile
(i.e. the field of the virtual photon) 
and the field of the emitted (real) photon.

With the effect of retardation accounted for, the field of the virtual 
photon is characterized by the 4-potential, $A_{\nu}$ ($\nu=0,1,2,3$),
whose components are given by
\begin{eqnarray}
A_{\nu}
=
\Q
\sum_{a=1}^{N}
\int
\d \bfr\,
\Psi_{\bfp_2\mu_2}^{(-){\hc}}(\bfr)\,
\gamma^{\mu}\,
D_{\mu\nu}(\omega,\bfr-\bfr_a)\,
\Psi_{\bfp_1\mu_1}^{(+)}(\bfr)\,.
\label{field.2}
\end{eqnarray}
Here $\Psi^{(\pm)}_{\nu}(\bfr)$ are the bi-spinor wavefunctions
corresponding to the out- (the upper index `$+$') and to the in- 
(`$-$') scattering states of the projectile,
the symbol $\hc$  denotes the hermitian conjugation,
$\gamma^{\mu}$ ($\mu=0,1,2,3$) are the Dirac matrices.
The quantity ${D_{\mu\nu}(\omega,\bfr-\bfr_a)}$ stands for the photon
propagator. The sum is carried out over the target electrons,
$\bfr_a$ is the coordinate of the $a$th electron.
 
The scalar part of the 4-vector (\ref{field.2}) $A_{0}\equiv \Phi$ 
defines the non-retarded Coulomb part of the interaction,
the spatial components of  $A_{\nu}$ ($\nu=1,2,3$) define the 
retarded vector potential $\bfA$ created by the projectile.
  
The consistent treatment of the relativistic effects in the PBrS
process implies that it is necessary to retain all multipoles 
when describing  the 
the vector potential $\bfA_{\gamma}$ of the emitted real photon:
\begin{eqnarray}
\bfA_{\gamma}
=
\sum_{a=1}^{N}
\bfe\,{\ee}^{-\i\bfk\bfr_a}\,.
\label{field.1}
\end{eqnarray}

Within the framework of perturbation theory the amplitude of PBrS
is given by the sum of two second-order matrix elements which 
correspond to the atom's transition 
$0 \rightarrow n \rightarrow 0$ under the action of
the fields $\bfA_{\gamma}$  and $A_{\nu}$:
\begin{eqnarray}
f_{\pol}
=
-e^2\,
\sum_{n}
\left\{
{
\langle 0| \bgamma\bfA_{\gamma} |n \rangle
\langle n| \gamma^{\nu}A_{\nu} | 0 \rangle
\over \E_n(1-{\rm i} 0) - \E_0 - \om }
+
{
\langle 0| \gamma^{\nu}A_{\nu} | n \rangle
\langle n| \bgamma\bfA_{\gamma}|0 \rangle
\over \E_n(1-{\rm i} 0) - \E_0 + \om }
\right\}
\label{PolAmpl.1}
\end{eqnarray}
The sum is carried out
over the quantum numbers of the complete spectrum of the (virtual) excited
atomic states and contains the contributions of 
the positive-energy, ${\E_n>0}$, and the negative-energy, ${\E_n<0}$, states.

The structure of the right-hand side of (\ref{PolAmpl.1}) is
similar to the non-realtivistic case (see Eq. (13) in \cite{NonRelBrS_2004}).
The difference is that in (\ref{PolAmpl.1}) all the quantities 
which refer to the particles are treated relativistically.

\subsection{The limiting cases of the amplitude (\protect\ref{PolAmpl.1}).}
\label{limits}

Let us demonstrate how the expressions for $f_{\pol}$, which one can
derive within the frameworks of simpler theories, follow from  
the general relativistic formula (\ref{PolAmpl.1}).

The {\em non-relativistic dipole-photon limit} of (\ref{PolAmpl.1}) 
one can obtain using the following transformations.
Firstly, carrying out the non-relativistic limit with respect to
the projectile, one notices that the vector potential 
$\bfA$ (which is due to the interaction retardation) can be neglected because 
of the relation $|\bfA|/|\Phi| \sim v_{1,2}\ll 1$.
Therefore, only the Coulomb component the 4-potential $A_{\nu}$ 
survives, where one uses $D_{00}(\omega,\bfr-\bfr_a)=1/|\bfr-\bfr_a|$.
Secondly, carrying out the limit $k\to 0$ in (\ref{field.1}) one
makes a substitution 
$\bgamma\bfA_{\gamma} \to \bfe\sum_{a=1}^{N}\hat{\bfp}_a\equiv  \bfe\hat{\bfP}$
where $\hat{\bfP}$ is the momentum operator of all atomic electrons.
The third step is to substitute $\gamma^{\nu}A_{\nu}$ with its
non-relativistic analogue equal to
$\Q\sum_{a}\langle\bfp_2^{(-)}\left|1/|\bfr-\bfr_a|\right|\bfp_1^{(+)}\rangle$.
As a result of these transformations the sum over the negative-energy
continuum in (\ref{PolAmpl.1}) becomes identically equal to zero, 
whereas the positive-energy sum reduces to the non-relativistic
limit of $f_{\pol}$ (see Eq. (13) in \cite{NonRelBrS_2004}).

To obtain the PBrS amplitude 
within the framework of {\em relativistic plane-wave first Born approximation}
(RBA) one substitutes the initial and  the 
final wavefunctions of the projectile 
with the field-free Dirac bi-spinors (see, e.g., \cite{Land4})
\begin{equation}
\Psi_{\bfp\mu}^{(\pm)}(\bfr)
=
u_{\mu}(\E,\bfp)\, \ee^{\i \bfp\bfr}\,
\label{RBA1.1}
\end{equation}
where $u_{\mu}(\E,\bfp)$ is the field-free bi-spinor amplitude,
and arrives at the expression derived in
\cite{OurRelativisticJPB,OurRelativisticJETP}.

If, within the RBA,  one carries out {\em the non-relativistic limit with
respect to the target's states}, and, additionally, {\em the dipole-photon
limit} $k\to 0$, the resulting formula reduces to the expression which 
was obtained for the first time in \cite{AmusiaKorolKuchievSolovyov1985}: 
\begin{eqnarray}
f_{\pol}
=
-4\pi\,\om
\Q
\left[
b^0\,
{\bfe\bfq \over q^2} \,
\alpha(\omega,q)
-
{\bfe\bfR \over \omega^2 - q^2} \,
\beta(\omega,q)
\right]\, .
\label{1_11b}
\end{eqnarray}
Here $\bfq = \bfp_1-\bfp_2$ is the momentum transfer,
$\bfR = \bfb - \bfq\,(\bfb\bfq)/q^{2}$ and the quantities 
$b^0$ and  $\bfb$ constitute a four-vector 
$b^{\mu}=\bar{u}_{\mu_2}(\E_2, \bfp_2)\,\gamma^{\nu}\,u_{\mu_1}(\E_1, \bfp_1)$.
The first term in the brackets contains the non-relativistic 
generalized polarizability $\alpha(\omega,q)$ of the target. 
This quantity enters the formula for the non-relativistic
PBrS amplitude, where it defines the dynamic response of the target to the
action of two fields: the  Coulomb field of the projectile and 
the field of the dipole photon (see \cite{NonRelBrS_2004}, Eq. (1)). 
The second term in the brackets is due to the retardation of the interaction
between the projectile and the target. 
It is proportional to another non-relativistic polarizability, 
$\beta(\omega,q)$, which is responsible for the dynamic response of the target
to the action of the retarded vector potential $\bfA$ and the 
dipole-photon field.
Explicit expressions for $\alpha(\omega,q)$ 
and $\beta(\omega,q)$ can be found in 
\cite{AmusiaKorolKuchievSolovyov1985,OurRelativisticJETP,OurRelativisticJPB}.

From formula (\ref{1_11b}) one easily derives the PBrS amplitude 
written within the frameworks of the {\em non-relativistic Born approximation and the 
dipole-photon approximation}.
Carrying out the limit $v_{1,2}\to 0$ one obtains
$b^0=1$ and $\bfb={\bf 0}$.
This results in $f_{\pol}=-4\pi\,\Q\,(\bfe\bfq)\,\om\,\alpha(\omega,q)/q^2$, 
which is a well-known result in the non-relativistic theory of PBrS.

Finally, let us briefly discuss the {\em high-energy photon limit} of 
the PBrS amplitude, 
i.e. when the photon energy exceeds the magnitude of the 
$K$-shell ionization potential $\om \gg I_{1s}$ \cite{OurRelativisticJPB}.
In the non-relativistic dipole-photon theory  this limit
is called the `striping' approximation 
\cite{AmusiaAvdoninaChernyshevaKuchiev1985,BuimistrovTrakhtenberg1977}.
In  \cite{OurRelativisticJPB} this limit was considered within 
the framework of the following approximation:  
relativistic description of the
projectile but non-relativistic treatment of the target.
The effects of the retardation and the emission into
higher multipoles are taken into account.

Similar to the non-relativistic dipole-photon case \cite{Korol1992}
the PBrS amplitude can be represented as a matrix element of the 
effective operator $V_{\rm eff}$ calculated between
the scattering states of the projectile.
Omitting the details we present the final result 
for $f_{\pol}$ in this limit:
\begin{eqnarray}
f_{\pol}
=
\langle \bfp_2 \mu_2 \left|V_{\rm eff}(\bfr) \right| \bfp_1\mu_1\rangle
\label{hep36a.1}
\end{eqnarray}
where the operator of $V_{\rm eff}(\bfr)$ is given by
\begin{eqnarray}
\fl
V_{\rm eff}(\bfr)
=
\i\,{e^2 \over m \om^2}
\int \d \bfr^{\prime}\,
\rho(r^{\prime})\,
\ee^{-\i \bfk\bfr^{\prime}}
\left[
\gamma_0\,{\bfe(\bfr-\bfr^{\prime}) \over \left|\bfr-\bfr^{\prime}\right|}
\left(
k
+
{\i \over \left|\bfr-\bfr^{\prime}\right|}
\right)
+
k\,\bfe\bgamma
\right]
{\ee^{\i k \left|\bfr-\bfr^{\prime}\right|}
\over\left|\bfr-\bfr^{\prime}\right| } .
\label{hep36a.6}
\end{eqnarray}
Here  $\rho(r^{\prime})$ is the density of the electron cloud of 
the target atom.

These formulae generalize the result of the non-relativistic dipole photon 
treatment of the PBrS process within the frame of the `stripping' 
approximation \cite{Korol1992,KorolLyalinSolovyovAvdoninaPratt2002}.
Indeed, carrying out the limit $k=0$ in (\ref{hep36a.6}) and substituting
in (\ref{hep36a.1}) the relativistic wavefunctions with the non-relativistic
ones one obtains
$f_{\pol}=-\om^{-2}\langle \bfp_2 \left|\bfe\bfa_{\rm el}\right| \bfp_1\rangle$
where $\bfa_{el}$ is the acceleration of the projectile due to the
static field created by the atomic electrons 
(see Eq. (22) in \cite{NonRelBrS_2004}). 


\subsection{The relativistic DPWA and multipole series
 for $f_{\pol}$.}
\label{PolBrSampl}

To derive the partial-wave series and the multipole series of the 
amplitude (\ref{PolAmpl.1}) one can use the following procedure 
\cite{OurRelativisticJPB}.
To start with  one uses the relativistic DPWA series for the 
wavefunctions $\Psi_{\bfp\mu}^{(\pm)}(\bfr)$ (see, e.g., \cite{Akhiezer}):
\begin{eqnarray}
\fl
\qquad
\Psi_{\bfp\mu}^{(\pm)}(\bfr)
=
{4\pi \over p r} 
\sum_{j l m}
\left(\Om_{j l m}^{\hc}(\bfn_\bfp)\, \chi_{\mu}(\bfn_\bfp)\right)
\, \ee^{\pm \i\delta_{j l}(\E)}
\pmatrix{
g(r)\, \Om_{j l m}(\bfn_\bfr)  \cr
- \i f(r) \,(\bsigma\bfn_\bfr)\, \Om_{jlm}(\bfn_\bfr) \cr}\,.
\label{ord1.4a}
\end{eqnarray}
Here a general notation $\bfn_{\bfa}$ is used for the unit vector in the
direction of $\bfa$,
$\Om_{j l m}(\bfn_\bfp)$ and $\Om_{j l m}(\bfn_\bfr)$
are the spherical spinors defined as in 
\cite{VarshalovichMoskalevKhersonskii},
$\chi_{\mu}$ stands for a two-component spinor
corresponding to the  spin projection $\mu$,
the quantities $\delta_{jl}(\E)$ are the scattering phaseshifts,
$\bsigma$ is the Pauli matrix.
The functions $g(r)\equiv g_{\E jl}(r)$ and $f(r)\equiv f_{\E jl}(r)$
are, correspondingly, the large and the small components of the
relativistic radial wavefunction in the central field of the target.
The sum is carried out over the total momentum $j$, orbital momentum $l$
and the projection $m$ of the total momentum.

The next step is to introduce the 
multipole expansions of the
factors $\bfe\bgamma\, \exp(-\i \bfk \bfr)$
and $\exp(-\i \bfq \bfr)$ (the latter appears in $A_{\nu}$ 
from (\ref{field.2}))
in terms of the vector spherical harmonics
$\bfY_{lm}^{(\lambda)}(\bfn)$ ($\lambda=0,1$)
and the spherical harmonics
$Y_{lm}(\bfn)= \bfn\bfY_{lm}^{(-1)}(\bfn)$ 
\cite{VarshalovichMoskalevKhersonskii}.

Finally, assuming that the states $| 0\rangle$ and $| n\rangle$
of the target can be described in terms of single-electron 
wavefunctions (see, e.g., \cite{LindgrenMorrison}),
one converts the sums over the atomic states $n$  from (\ref{PolAmpl.1})
into the sums over the quantum numbers,
$\E_\i, j_\i, l_\i, m_\i$ ($\i=0,n$) of the core and the excited subshells.
The bi-spinor single-electron wavefunctions,
$\Psi_{\E_\i j_\i l_\i m_\i}(\bfr)$, 
can be obtained by solving the system of self-consistent radial
Hartree-Fock-Dirac equations (see, e.g., \cite{ChernyshevaYakhontov1999}).

Having done all this and carrying out the intermediate algebra,
one obtains the following expression for 
$f_{\pol}$ written in terms of the multipole
series over $\bfY_{lm}^{(\lambda)}(\bfn_\bfk)$ and the
DPWA series with respect to the initial and the final states of 
the projectile:
\begin{eqnarray}
\fl
f_{\pol}
&=
\i\,
{(4\pi)^{5/2}\, \Q \over  p_1 p_2}
\!\! \sum_{j_1 l_1 m_1\atop j_2 l_2 m_2}
\sum_{\lambda=0,1\atop lm}
(-1)^{m_1+{1\over 2}}\,
\i^{-l-\lambda}\,
\ee^{\i\delta_{j_1 l_1}(\E_1) +\i\delta_{j_2 l_2}(\E_2)}
\left( \chi_{\mu_2}^{\hc}\Om_{j_2 l_2 m_2}\right)
\left(\Om_{j_1 l_1 m_1}^{\hc} \chi_{\mu_1}\right)
\nonumber \\
\fl
&
\times
\xi(l_2 l_1 l \lambda 1)\,
\Pi_{j_1j_2l}
\threejot{\! j_2}{\!\!j_1}{\!\!\! l}
         {\!-m_2}{\!\!m_1}{\!\!\!-m}
\threejot{ j_2}       {\!j_1}        {\!l}
          {{1\over 2}}{\!-{1\over 2}}{\!0}
\bfe\bfY_{lm}^{(\lambda)}(\bfn_\bfk)\,
\cP_{21}^{(\lambda)}(\omega,k,l)\, ,
\label{summary2.1}
\end{eqnarray}
where 
$\threejot{a}{b}{c}{\alpha}{\beta}{\gamma}$ stands for the $3j$-symbol, 
and the short-hand notation
$\Pi_{j_1j_2\dots}=\sqrt{(2j_1+1)(2j_2+1)\dots}$ is used
\cite{VarshalovichMoskalevKhersonskii}.
The function $\xi(l_2 l_1 l \lambda 1)$ equals to one 
if the sum of its
arguments is even, and equals to zero if otherwise.

The quantities ${\cP_{21}^{(\lambda)}(\om,k,l)}$ ($\lambda=0,1$) are
the relativistic partial amplitudes of the PBrS.
They depend, apart from the quantum numbers $\om,k,l$ on the combination
of the types of the real and the virtual photons entangled in the amplitude.
For a spherically-symmetric target there are three allowed combinations
of virtual and real photons: 
longitudinal-electric, electric-electric and magnetic-magnetic.
The first two combinations are incorporated in the amplitude
${\cP_{21}^{(1)}(\om,k,l)}$ which  consists  of two terms,
$\cP_{21}^{(l)}(\om,k,l)+\cP_{21}^{(e)}(\om,k,l)$.
The partial amplitude ${\cP_{21}^{(0)}(\omega,k,l)}$
corresponds to the magnetic-magnetic combination.
The explicit expressions for these terms are as follows:
\begin{eqnarray}
\cP_{21}^{(l)}(\om,k,l)
&=
-
{2 \over \pi}\,{\sqrt{l(l+1)}\over 2l+1}
\int_0^{\infty} q\, \d q\,
\alpha_{l}(\om,q,k)\,f_{21}^{(-1)}(q;l)
\label{J_s}\\
\cP_{21}^{(e)}(\om,k,l)
&=
{2 \over \pi}\,{\left[l(l+1)\right]^{3/2}\over 2l+1}
\int_0^{\infty}
{q^2\, \d q\ \over k^2- q^2 +\i 0}\,
 \beta_l^{(1)}(\om,q,k)\,f_{21}^{(1)}(q;l)
\label{J_e}\\
\bs
\cP_{21}^{(0)}(\om,k,l)
&=
- {2 \over \pi}\,{\sqrt{l(l+1)}\over 2l+1}
\int_0^{\infty}
{q^2\, \d q\ \over k^2- q^2+\i 0 }\,
\beta_l^{(0)}(\om,q,k)\, f_{21}^{(0)}(q;l)\, .
\label{J_m}
\end{eqnarray}
Here ${f_{21}^{(-1,0,1)}(q;l)}$ stand for the radial integrals 
which contain the radial wavefunctions (the large and small components)
of the initial and the final states of the projectile and
the spherical Bessel functions $j_l(qr)$.
The corresponding formulae can be found in \cite{OurRelativisticJPB}.

The most important feature of the formulae (\ref{J_s})--(\ref{J_m})
is that they clearly demonstrate that in the relativistic case
the PBrS amplitude is expressed  in terms of the multipole generalized 
dynamic polarizabilities of 
three different types corresponding to the allowed combinations of the 
types of the virtual/real photons: 
(a) $\alpha_l(\om,q,k)$ corresponds to the longitudinal/electric 
combination, 
(b) $\beta_l^{(1)}(\om,q,k)$ to the  electric/electric one,
and
(c) $\beta_l^{(0)}(\om,q,k)$ to the magnetic/magnetic.
Each of these polarizabilities depends on the
photon energy $\om$, its orbital momentum $l$
(which defines the multipolarity of the polarizability), 
and on the magnitudes of the momenta $q$ and $k$ 
of the virtual and real photons.
We do not reproduce here the explicit expressions for the polarizabilities
but refer to the papers \cite{OurRelativisticJPB,OurRelativisticJETP} where
various representations are given and discussed in detail. 

In the non-relativistic limit (with respect to both the projectile
and to the target) and in the {\em dipole-photon} approximation ($k=0$) 
the amplitudes $\cP_{21}^{(e)}(\om,k,l)$ and  $\cP_{21}^{(0)}(\om,k,l)$
vanish for all $l$. 
The only term which survives is $\cP_{21}^{(l)}(\om,k,l)$
with $l=1$.
The corresponding generalized polarizability $\alpha_1(\om,q,0)$ reduces
to the non-relativistic polarizability $\alpha(\om,q)$, and 
the partial amplitude $\cP_{21}^{(1)}(\om,0,l)$ reproduces the 
non-relativistic expression (see Eq. (18) in \cite{NonRelBrS_2004}).

Let us note that the polarizability $\beta(\om,q)$ introduced
above in Eq. (\ref{1_11b}) is equal to $\beta_1^{(1)}(\om,q,0)$.

Finally, we mention that the structure of the DPWA and multipole
series for the OBrS amplitude is similar to that of the right-hand side
of (\ref{summary2.1}).
Therefore, to obtain the total BrS amplitude,
$f_{\tot}=f_{\ord}+f_{\pol}$ one substitutes $\cP_{21}^{(\lambda)}(\omega,k,l)$
in (\ref{summary2.1}) with 
$\cT_{21}^{(\lambda)}(\omega,k,l)=
\cP_{21}^{(\lambda)}(\omega,k,l)+\cO_{21}^{(\lambda)}(\omega, k,l)$, where
 $\cO_{21}^{(\lambda)}(\omega,k,l)$ are the OBrS partial amplitudes
\cite{OurRelativisticJPB}.
In the non-relativistic limit, carried out with respect to the projectile
motion and the states of atomic electrons, and in the 
dipole-photon limit the DPWA series (\ref{summary2.1}) 
reproduce the formulae presented 
in \cite{AmusiaKorol1992}.

\subsection{The cross section of relativistic PBrS}
\label{RelBrS_CS}

The expansion (\ref{summary2.1}) allows one to derive
the DPWA and multipole series for the spectral and 
spectral-angular distributions of PBrS (and the total BrS as well).
The details of the formalism and the final results are 
presented in \cite{OurRelativisticJPB}.

Aiming to point out the qualitative differences, 
which are model-independent, between the relativistic case and 
the non-relativistic one we discuss the PBrS cross section 
in terms of the relativistic plane-wave Born approximation.

We start with the discussion of the spectral distribution of PBrS,
$\d\sigma_{\pol}/\d\om$.
The corresponding cross section 
$\d\sigma_{\pol}(\om)=\om\,(\d\sigma_{\pol}/\d\om)$ is given by
\cite{OurRelativisticJETP}:
\begin{eqnarray}
\fl
\d\sigma_{\pol}(\om)
&=
 \alpha \,{2 Z_{\rp}^2\,\om^2 \over  p_1^2}
\sum_{l=1}^{\infty}
{ l (l+1) \over 2l+1}\,
\int\limits_{\qmin}^{\qmax} {\d q \over q}\,
\Biggr[
\Bigr(q_{\min}^2\, q_{\max}^2 - q^2k^2\Bigl)
\left|\alpha_{l}(\omega,q,k) \right|^2
\nonumber\\
\bs
\fl
&+
l (l+1)\,
{2q^2(q^2 -k^2) +(q^2-q_{\min}^2)(q_{\max}^2 -q^2)
\over 2(q^2 -k^2)^2}
\sum_{\lambda=0,1}
\left|\beta_{l}^{(\lambda)}(\omega,q,k)\right|^2
\Biggr]
\label{RBA.spectr}
\end{eqnarray}
Here $\alpha$ is the fine structure constant, 
$q = \left|\bfp_1 - \bfp_2\right|$ is the momentum transfer.
Its minimum and maximum values are $\qmin=p_1-p_2$ and $\qmax=p_1+p_2$.

The first term in the integrand, proportional to 
$\left|\alpha_{l}(\omega,q,k) \right|^2$, is due to the dynamic polarization
of the target by the Coulombic part of the field of the projectile,
whereas the second term, containing the polarizabilities
$\beta_{l}^{(\lambda)}(\omega,q,k)$  appears as the result of the 
retardation of the interaction.
In the non-relativistic limit (with respect to the projectile and the target)
and in the dipole-photon regime all polarizabilities but 
$\alpha_1(\omega,q,k)$ are equal to zero. 
In this limit $\alpha_1(\omega,q,k) \to \alpha(\omega,q)$, which is the
non-relativistic dipole generalized polarizability, and the right-hand
side of (\ref{RBA.spectr}) reduces to the non-relativistic
PBrS cross section (see Eq. (24) in \cite{NonRelBrS_2004}).
Similar to the non-relativistic result the cross section (\ref{RBA.spectr})
weakly depends on the mass of a projectile.

An important distinguishing feature of the {\em relativistic} PBrS cross section 
is its logarithmic growth with  $\E_1$ 
\cite{AmusiaKorolKuchievSolovyov1985,
AstapenkoBuimistrovKrotovMikhailovTrahtenberg1985}.
Qualitatively, the reason for this is as follows.
As was already mentioned, unlike a non-relativistic particle, a relativistic
one interacts with the target not only by its Coulomb field but also
(in the ultra-relativistic case, predominantly) by the field of transverse
virtual photons.
The effective radius of this field, $R_{\rm eff}$, increases infinitely
with the energy of the projectile. 
The magnitude of $R_{\rm eff}$ can be estimated as 
$R_{\rm eff} \sim 1/q_{\min}^{\perp}$, where
$q_{\min}^{\perp}\approx \gamma^{-1}\qmin$, with
$\gamma=\E_1/m$ and $q_{\min}=p_1-p_2\approx \om/v_1$
\cite{AmusiaKorolKuchievSolovyov1985}. 
As a result, the range of distances at which a projectile effectively 
polarizes a target increases. 
This leads to the growth of $\d\sigma_{\pol}(\om)$ with $\E_1$.

Quantitative description of this effect can be carried out
directly from (\ref{RBA.spectr}) by analyzing the
contribution of the region $q_{\rm at} \geq q \geq \qmin\approx \om/v_1$ 
to the  integral (here $q_{\rm at} = R_{\rm at}^{-1}$ with 
$R_{\rm at}$ being the radius of the target).
The result reads \cite{OurRelativisticJETP}:
\begin{eqnarray}
\Bigl[\d\sigma_{\pol}(\om)\Bigr]_{q\sim \qmin}
\approx
 \alpha \,{4 Z_{\rp}^2\,\om^4 \over  v_1^2}
\left(
A\, \ln{q_{\rm at} \over \qmin}
+
B \, \ln{\E_1 \over m}
\right)
\label{RBA.45}
\end{eqnarray}
where 
\begin{eqnarray}
\fl
A =
\sum_{l=1}^{\infty}
{ 2l (l+1) \over 2l+1}\left|\alpha_{l}(\omega,\qmin,k) \right|^2,
\quad
B =
{1 \over \om^4}
\sum_{l=1}^{\infty}
{ l^2 (l+1)^2 \over 2l+1}
\sum_{\lambda=0,1}
\left|\beta_{l}^{(\lambda)}(\omega,\qmin,k)\right|^2\, .
\label{RBA.45b}
\end{eqnarray}
The first term in term in the brackets on the right-hand side of (\ref{RBA.45})
corresponds to the contribution of the Coulomb part of the interaction, and
it remains constant as the energy $\E_1$ grows.
The second term, which is due to the interaction retardation, 
logarithmically increases with $\E_1$.
This feature was first noted in 
\cite{AmusiaKorolKuchievSolovyov1985,
AstapenkoBuimistrovKrotovMikhailovTrahtenberg1985}, where
the PBrS of a relativistic projectile was studied 
within the framework of the non-relativistic dipole-photon
description of the target.
The result of these studies follows from more general
expressions (\ref{RBA.45}) and (\ref{RBA.45b}).
Carrying out the non-relativistic dipole photon limit, and
accounting for the relations 
$\lim_{k\to 0}\alpha_{1}(\omega,\qmin,k)
=\lim_{k\to 0}\om^{-2}\beta_{1}^{(1)}(\omega,\qmin,k)
\approx \alpha_{\rm d}(\om)$ if $\qmin R_{\rm at}\ll 1$
one obtains \cite{AmusiaKorolKuchievSolovyov1985}:
\begin{eqnarray}
\d\sigma_{\pol}(\om)
\approx
 \alpha \,{16 Z_{\rp}^2\,\om^4 \over 3 v_1^2}\,
\om^4 \left|\alpha_{\d}(\omega) \right|^2\, 
\ln{q_{\rm at}\E_1 \over m\qmin}\,,
\label{RBA.45c}
\end{eqnarray}
where $\alpha_{\d}(\omega)$ is the dynamic (non-relativistic) dipole 
polarizability.

\begin{figure}
\begin{center}
\hspace{-5.5mm}
\includegraphics[scale=0.3,angle=270]{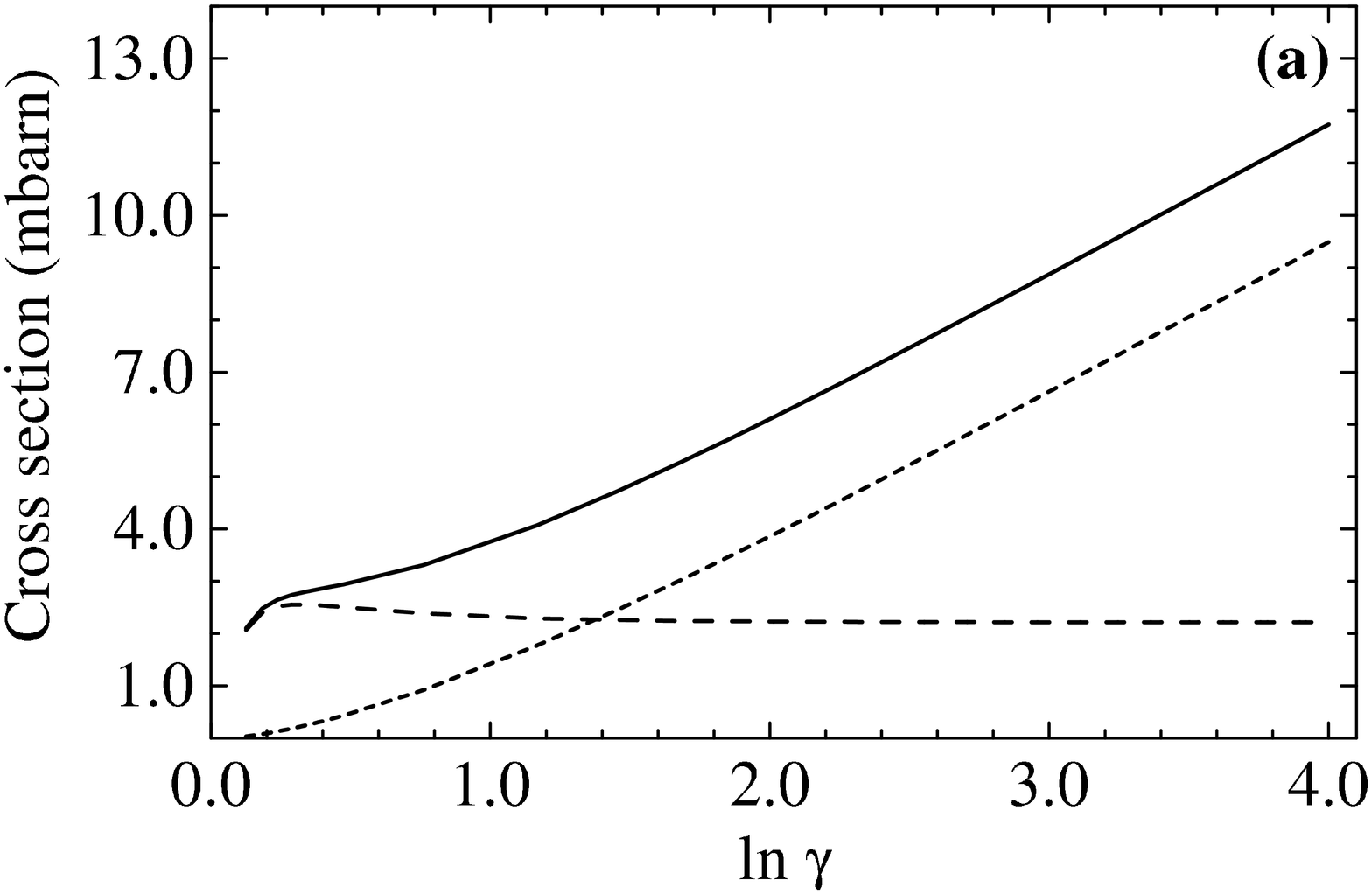}
\quad
\includegraphics[scale=0.3,angle=270]{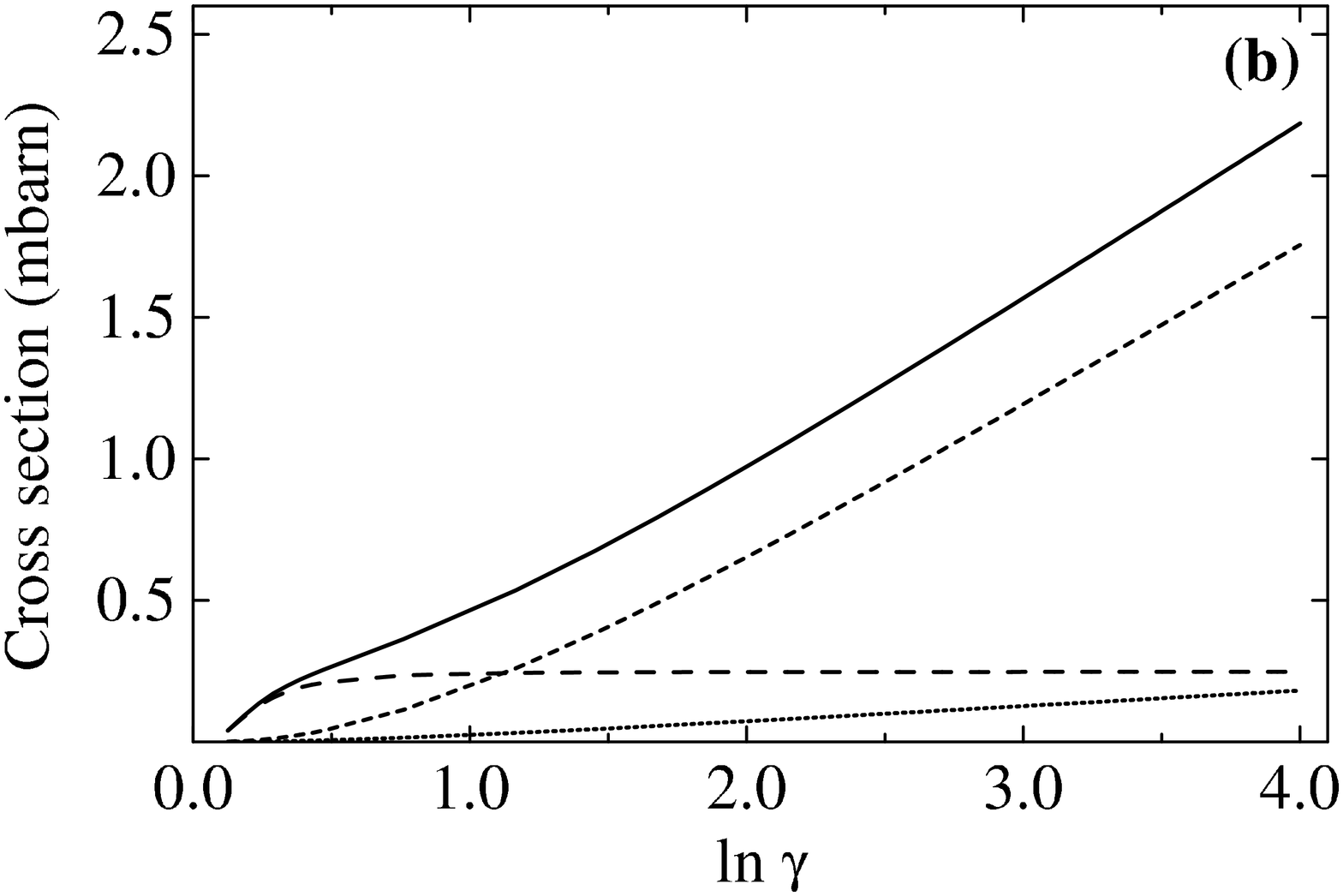}
\end{center}                                                     
\caption{Dependence of $\d\sigma_{\pol}(\om)$ 
on $\ln\gamma \equiv \ln(\E_1/m)$ for a proton-Au$^{+78}$ collision.
Thick solid curve describes $\d\sigma_{\pol}(\om)$ with all terms
on the right-hand side of (\protect\ref{RBA.spectr}) included.
The long-dashed, the dotted and the short-dashed curves represent,
respectively, the contributions of the terms containing 
the polarizabilities $\alpha_{l}(\omega,q,k)$,
$\beta_{l}^{(0)}(\omega,q,k)$ and $\beta_{l}^{(1)}(\omega,q,k)$.
In figure (a) the photon energy $\om$ equals to $1.5 I$, in figure
(b) $\om=4I$ (where $I=93.5$ keV is the ionization potential of 
the target).
 }
\label{fig_E-dep}
\end{figure} 
The logarithmic growth of $\d\sigma_{\pol}(\om)$ with $\E_1$
is illustrated  in figures \ref{fig_E-dep}a,b where the PBrS
cross sections formed in the collision of a proton with
a hydrogen-like gold are presented.
Because of the large mass of the proton its OBrS
is suppressed by the factor $m^{-2}\sim 10^{-6}$, and can be
neglected.
The data presented in the figures correspond to two values of $\om$ 
as indicated in the caption.
The growth of the cross section, which is nearly linear for 
high values of $\ln(\E_1/m) $ in accordance with (\ref{RBA.45}) and 
(\ref{RBA.45c}),
is due to the contribution  of the terms containing 
$\beta_{l}^{(0)}(\omega,q,k)$ and $\beta_{l}^{(1)}(\omega,q,k)$
(see (\ref{RBA.spectr})).
The contribution of the terms proportional to $\alpha_{l}(\omega,q,k)$
is virtually independent on $\E_1$.

Expression (\ref{RBA.45c}) generalizes the result of non-relativistic theory.
Indeed, putting $\E_1=m$ on the right-hand side one obtains the
expression derived in \cite{Zon1977,AmusiaZimkinaKuchiev1982}, where it
was called  the `logarithmic approximation' for PBrS.
Let us note here that the OBrS cross section of a relativistic 
projectile in collision with a neutral target is almost independent
on $\E_1$ for high velocities of the projectile, $v_1\leq 1$.
Therefore, the total cross section also increases logarithmically with $\E_1$.

In the case of an ultra-relativistic electron or positron 
($m_e=Z_{\rp}^2=1$) scattering on a neutral target there is a peculiar 
feature of the total BrS cross section which is clearly distinct 
from the non-relativistic case.
To demonstrate it we consider the limit of high energy photons
when $\om \gg I_{1s}$, and one can use the approximate formula for
the dipole polarizability $\alpha_{\d}(\omega)\approx -Z/\om^2$
($Z$ stands for the number of atomic electrons which for a neutral
atom coincides with the atomic number).
The PBrS cross section (\ref{RBA.45c}) reduces to
$\d\sigma_{\pol}(\om)\approx  {16\alpha /3}\,Z^2\ln(q_{\rm at}\E_1/\om)$
where we used $v_1 =1$  and $\qmin = \om$.
In the `logarithmic approximation' the OBrS cross section is given by
(see, e.g., \cite{Land4}) 
$\d\sigma_{\ord}(\om)\approx  {16\alpha /3}\,Z^2\ln(1/q_{\rm at})$.
In the same approximation one can neglect 
the interference term and derive the following
expression for the total BrS cross section:
\begin{eqnarray}
\d\sigma_{\tot}(\om)
\approx
\d\sigma_{\ord}(\om)
+
\d\sigma_{\pol}(\om)
\approx
 Z^2 \,{16 \alpha\over 3}\,\ln{\E_1 \over \om}\,.
\label{RBA.45d}
\end{eqnarray}
Apart from the factor $Z^2$ the right-hand side of this equation
reproduces the formula for the cross section of BrS emitted by 
a slow free electron in the collision with an ultra-relativistic
electron or positron \cite{Land4}.
This coincidence is not accidental and has clear qualitative explanation
\cite{AmusiaKorolKuchievSolovyov1985}.
For $\om \gg I_{1s}$ the atomic electrons can be treated as free ones.
If $Z\alpha \ll 1$ then the velocities of all atomic electrons are small
compared to that of the projectile.
Then, the total BrS amplitude can be written as the sum of three terms
$f_{\tot} =Z \cF_1 + Z\cF_2 + \cF_3$.
Here $\cF_1$ denotes the amplitude of the photon emission by the projectile
electron/positron interacting with a free atomic electron,
$\cF_2$ is the amplitude of the emission by the atomic electron during
this interaction, and
$\cF_3$ is the amplitude due to the interaction of the
projectile with the nucleus.  
As known (see, e.g., \cite{Land4}),
the BrS amplitude of an ultra-relativistic electron/positron scattered
from a free slow particle depends on the particle's charge and does not 
depend on its mass.
Therefore, the sum of two terms, $Z\cF_1 + \cF_3$ is identically equal
to zero, and $f_{\tot}$ reduces to $Z \cF_1$, 
i.e. only atomic electrons radiate in this process.
In a way, this result is opposite to the 'stripping' effect in the 
non-relativistic BrS, when $\d\sigma_{\tot}(\om)$ in 
electron-atom collision reduces to that 
on the bare nucleus if $\om \gg I_{1s}$
(see Eq. (6) in \cite{NonRelBrS_2004}). 
The qualitative explanation of this difference between the non-relativistic
and the ultra-relativistic cases is as follows.
In the former case, the range of distances between the projectile and the
target important in the PBrS process can be estimated 
as $R_{\rm eff} \sim 1/\qmin \approx v_1/\om$.
This value is much smaller than the photon wavelength $\lambda=2\pi/\om$.
Therefore, the dipole approximation can be applied 
for the system 'the projectile + the target'. 
The only allowed radiation in this system in the range $\om \gg I_{1s}$
(the limit of quasi-free atomic electrons) is that by a projectile on  
the nucleus.
As was already mentioned, in the ultra-relativistic case
$R_{\rm eff} \sim 1/q_{\min}^{\perp} \approx \gamma/\om$.
As $v_1\to 1$ and $\gamma \to \infty$ these distances increase
unrestrictedly and are much greater than $\lambda$. 
The retardation effects in the interaction between the projectile and the 
atomic electron become important, so that the dipole approximation is 
inapplicable to this system. 
This leads to the difference of (\ref{RBA.45d}) from its non-relativistic 
analogue.

The retardation, as well as the relativistic and higher multipoles effects,
strongly modify not only
the PBrS spectral distribution but the spectral-angular distribution 
as well \cite{OurRelativisticJETP}.
The spectral-angular distribution can be written in the form of series
over the Legendre polynomials $P_{l}(\cos\theta)$:
\begin{eqnarray}
\d^2 \sigma_{\pol}(\om,\Om)
\equiv
\om\,{\d^2 \sigma_{\pol} \over \d\om \d\Om}
=
{\d \sigma_{\pol}(\om) \over 4\pi}
\left(
1+
\sum_{l_k=1}^{\infty}\,
a_{l_k}(\om)\,P_{l_k}(\cos\theta)
\right)\,,
\label{RBA.117}
\end{eqnarray}
here $\theta$ is the emission angle (measures with respect to
$\bfp_1$).
The coefficients $a_{l_k}(\om)$ depend on the photon energy and 
are the bi-linear forms of the relativistic polarizabilities
$\alpha_{l}(\omega,q,k)$ and $\beta_{l}^{(1,2)}(\omega,q,k)$.
The explicit formulae for $a_{l_k}(\om)$, obtained within the DPWA
and the relativistic Born approximation, can be found in the 
cited paper (see also \cite{OurRelativisticJPB}).

The expression for $\d^2 \sigma_{\pol}^{\NR}(\om,\Om)$ obtained within the
framework of non-relativistic dipole-photon approximation
reads \cite{AmusiaAvdoninaKuchievChernysheva1986,AmusiaKorol1992}:
\begin{eqnarray}
\d^2 \sigma_{\pol}^{\NR}(\om,\Om)
=
{\d \sigma_{\pol}^{\NR}(\om) \over 4\pi}
\left(
1+
\beta(\om)\,P_{2}(\cos\theta)
\right)\,.
\label{RBA.117a}
\end{eqnarray}
The quantity $\beta(\om)$ is frequently 
called a (dipole) coefficient of angular anisotropy.

Let us point to the differences between (\ref{RBA.117}) and (\ref{RBA.117a}).
The first one, already discussed above, concerns the different 
dependence of the factors $\d \sigma_{\pol}(\om)$ and 
$\d \sigma_{\pol}^{\NR}(\om)$ on $\E_1$.
The former increases proportionally to $\ln(\E_1)$ and this is due to
the retardation.
The second difference, a rather obvious one, 
is reflected by an infinite number of terms on the right-hand side of 
(\ref{RBA.117}), which are due to
the infinite number of the photon multipoles taken into account,  
in contrast to the two terms in (\ref{RBA.117a}).
Finally, there is a `hidden' difference which is related to the 
relativistic description of the internal dynamics of the target 
rather than to the multipole character of the radiation.
The easiest way to trace the origin of this difference is to consider
the contribution of only dipole photons to the series  in (\ref{RBA.117}).
The momentum of a dipole photon is $l=1$, therefore only the amplitudes
$\cP_{21}^{(l,e,m)}(\om,k,1)$ (see (\ref{J_s})--(\ref{J_m})) 
will contribute to the PBrS amplitude (\ref{summary2.1}). 
Then, instead of (\ref{RBA.117}) one obtains
\cite{OurRelativisticJETP}:
\begin{eqnarray}
\fl
\Bigl[\d^2 \sigma_{\pol}(\om,\Om)\Bigr]_{l=1}
=
{\Bigl[\d \sigma_{\pol}(\om)\Bigr]_{l=1} \over 4\pi}
\left(
1
+
a_{1}(\om)\,P_{1}(\cos\theta)
+
a_{2}(\om)\,P_{2}(\cos\theta)
\right)\,,
\label{RBA.117b}
\end{eqnarray}
It is seen that in contrast to (\ref{RBA.117a}) the angular distribution
of the relativistic {\em dipole} radiation contains the term proportional to 
$P_{1}(\cos\theta)$.
As a result, the angular distribution becomes
asymmetric with respect to the transformation 
$\theta \longrightarrow \pi-\theta$.

The reason for this effect is as follows.
If the target is treated within the non-relativistic framework then
the dipole photons emitted via the polarizational mechanism
belong to the `electric' type  \cite{Land4}.
This statement is valid for arbitrary velocities of the projectile.
Indeed, for $v_1\ll 1$ the PBrS amplitude is proportional to
$\alpha(\om,q)$ (see Eq. (1) in \cite{NonRelBrS_2004}) which couples
the Coulomb field of a projectile and the field of the `electric'
dipole photon.
For relativistic velocities, $v_1 \leq 1$, the amplitude $f_{\pol}$
contains two polarizabilities, $\alpha(\om,q)$ and $\beta(\om,q)$
(see (\ref{1_11b})). 
The former couples the  `electric' dipole photon with the `electric'
dipole virtual photon.
Due to the selection rules, the angular dependent part of the intensity 
of the `electric' dipole radiation contains the term proportional
to $P_2(\cos\theta)$ only \cite{Land4}.
In the most explicit form this can be illustrated if one evaluates
the angular distribution of PBrS using the 'logarithmic approximation'.
Then, either for a non-relativistic projectile \cite{Zon1977} or a 
relativistic one \cite{AmusiaKorolKuchievSolovyov1985} 
the PBrS angular distribution is proportional to
$1+P_2(\cos\theta)/2\propto 1+ \cos^2\theta$, which coincides with the
 angular distribution of radiation emitted by
a rotating electric dipole \cite{Landau2}.

In the case when the internal dynamics of the target is treated
within the relativistic theory, there appears a possibility to emit a
photon of the  `magnetic' type.  
If the `magnetic' dipole radiation is treated separately, its 
angular dependent part also contains the term proportional
to $P_2(\cos\theta)$ only \cite{Land4}.
However, if both types of the dipole photons can be emitted in a
process, then their interference results in the term proportional
to $P_1(\cos\theta)$.
This is exactly what happens in the relativistic PBrS.
Using the general formulae presented in 
\cite{OurRelativisticJPB,OurRelativisticJETP} one can establish, that the 
coefficient $a_{1}(\om)$ in (\ref{RBA.117b}) is proportional
to the cross-terms containing the products of the relativistic polarizability
$\beta_{1}^{(0)}(\omega,q,k)$ of a `magnetic'-type 
with the polarizabilities $\beta_{1}^{(1)}(\omega,q,k)$ and
$\alpha_{1}(\omega,q,k)$ which belong to the 
`electric'-type.
The coefficient $a_{1}(\om)$ contains the quadratic forms of
the polarizabilities of the same type, i.e. the terms proportional to
$\left|\beta_{1}^{(\lambda)}(\omega,q,k)\right|^2$ (with $\lambda=0,1$) and
to $\alpha_{1}(\omega,q,k)\,\beta_{1}^{(1)}(\omega,q,k)$.
The magnitude of $a_{1}(\om)$ relative to $a_{2}(\om)$ 
is defined by the ratios
$\left|\beta_{1}^{(0)}(\omega,q,k)/\beta_{1}^{(1)}(\omega,q,k)\right|$
and 
$\left|\beta_{1}^{(0)}(\omega,q,k)/\alpha_{1}(\omega,q,k)\right|$
which increase with the photon energy $\om$ or/and the charge of the nucleus
$Z$, i.e. the factors responsible for the magnitude of the 
relativistic effects in atomic radiative processes.

\begin{figure}
\begin{center}
\hspace{-6.5mm}
\includegraphics[scale=0.29,angle=270]{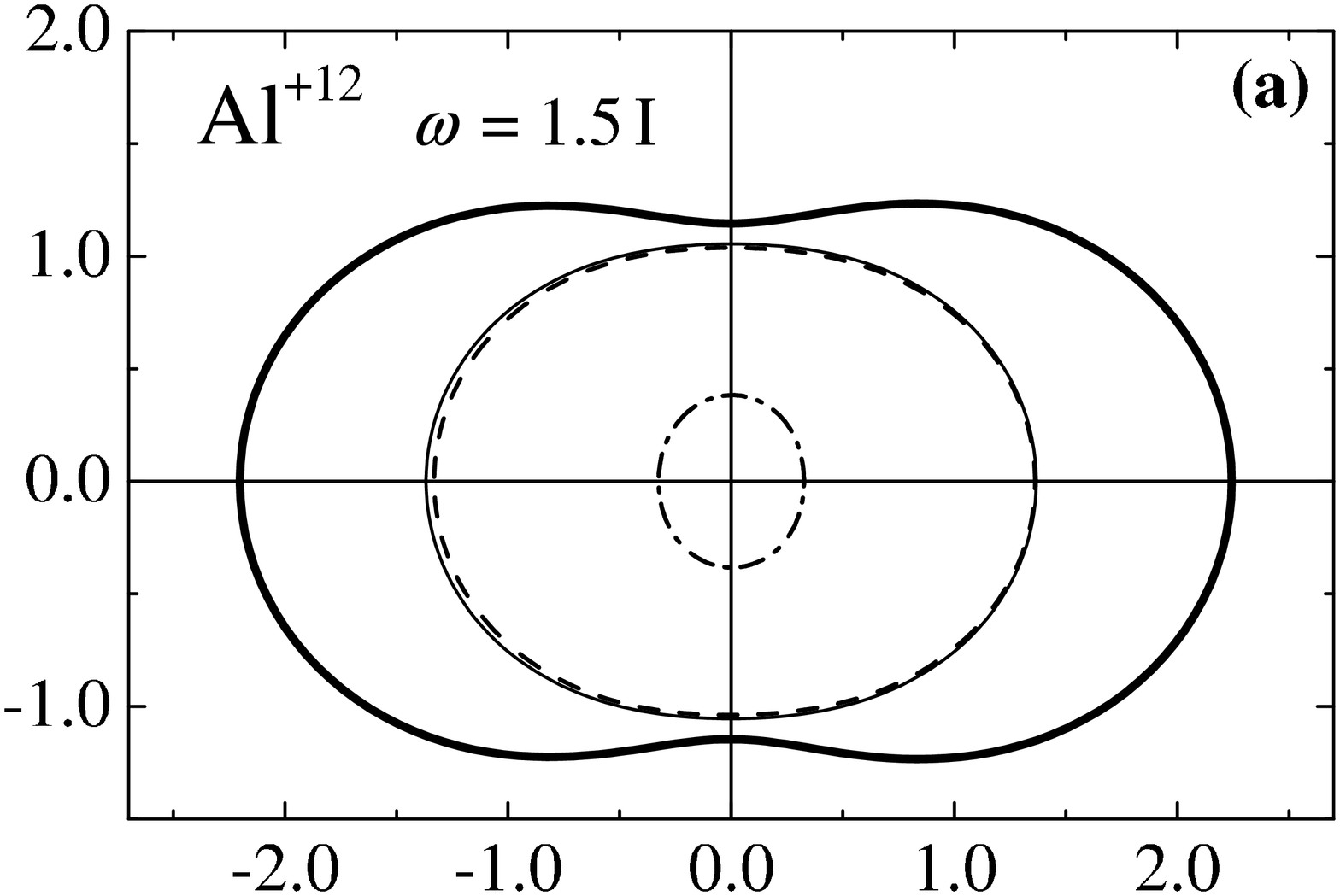}
\ 
\includegraphics[scale=0.29,angle=270]{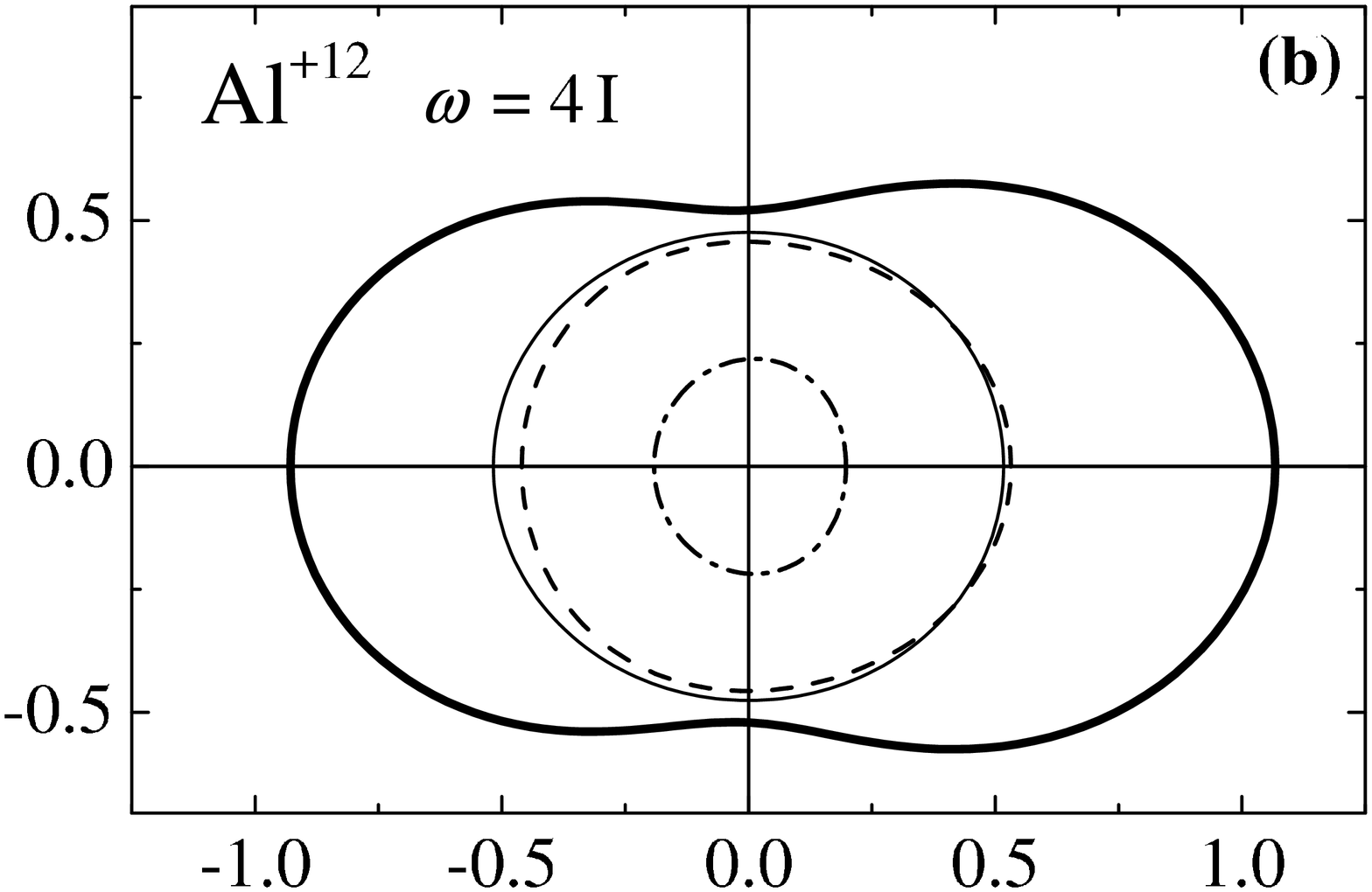}
\end{center}                                                     
\begin{center}
\includegraphics[scale=0.45,angle=0]{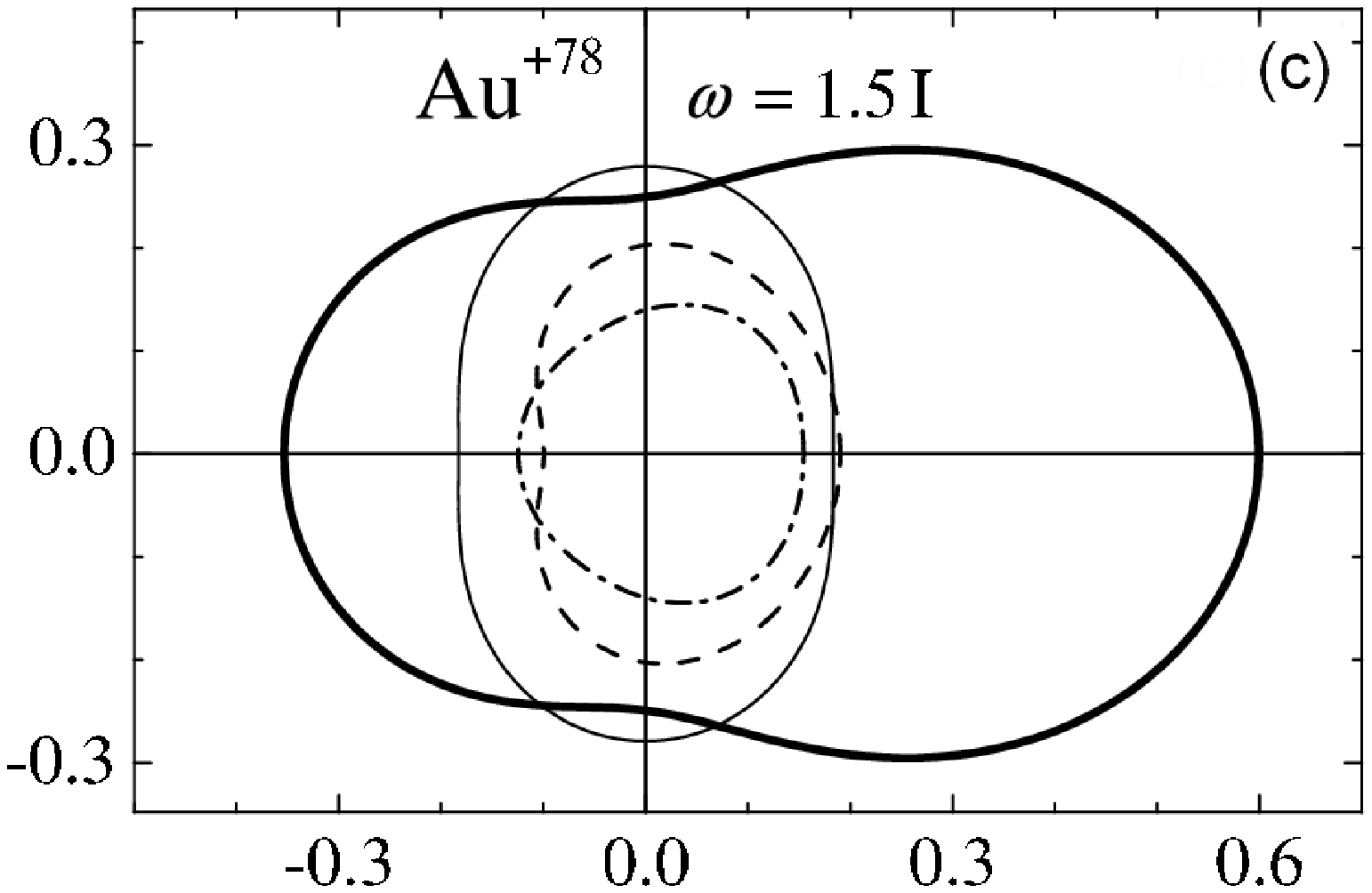}
\hspace*{0.8cm}
\includegraphics[scale=0.46,angle=0]{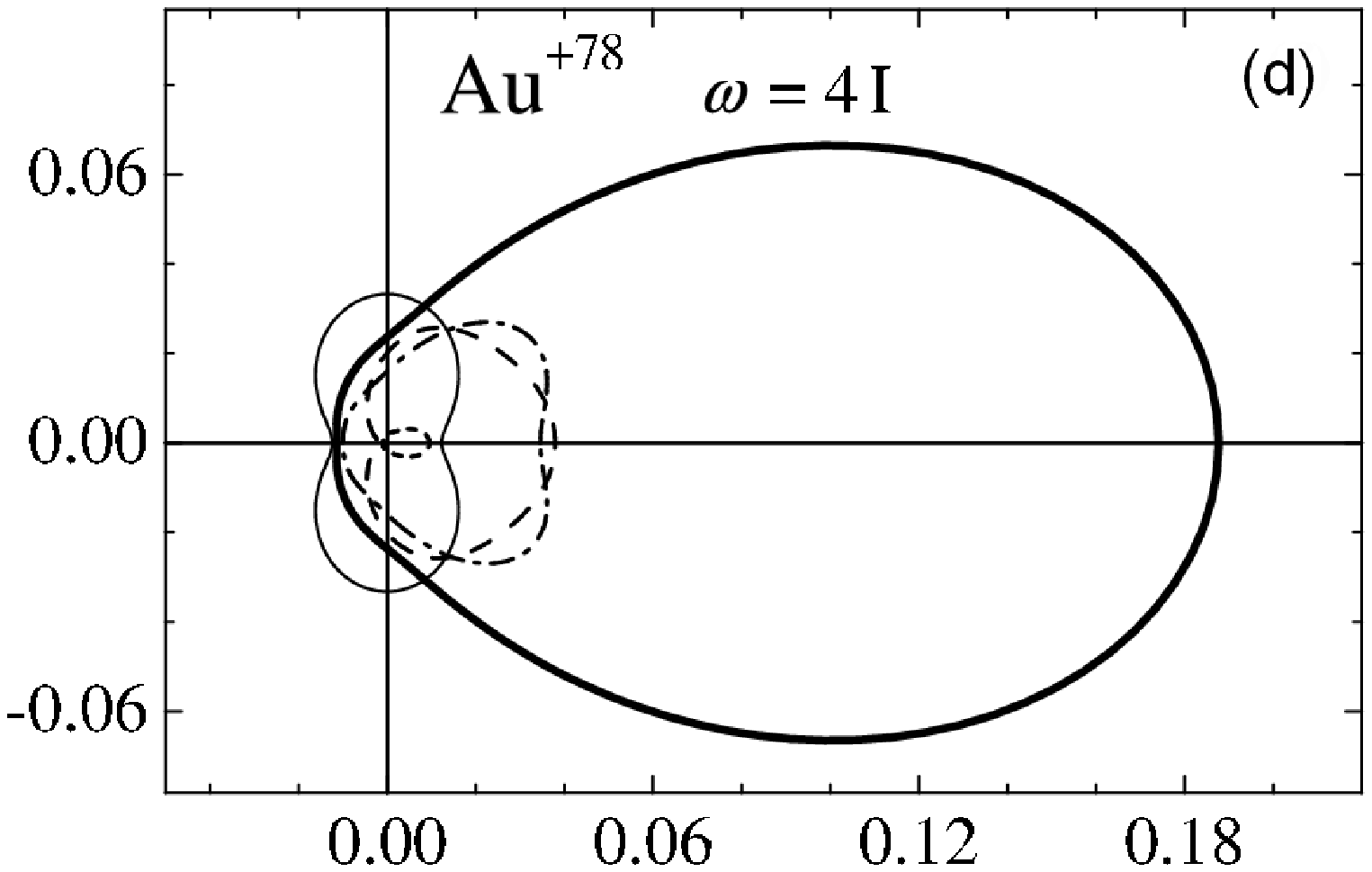}
\end{center}                                                     
\caption{Profiles of the angular distribution of PBrS 
$\d^2 \sigma_{\pol}(\om,\Om)$
for collisions of 3 GeV protons with Al$^{+12}$ (figures a and b), 
 and Au$^{+78}$ (c and d) ions calculated for two 
photon energies.
In figures a and c $\om=1.5 I$, in figures b and d $\om=4I$.
Thick solid curves stand for the fully relativistic calculations
(\protect\ref{RBA.117}), thin solid
curves correspond to the non-relativistic dipole-photon approximation
(\protect\ref{RBA.117a}).
The dashed, dotted, and dash-dotted  curves correspond to the contributions 
of the terms proportional to the squares of the moduli of 
the polarizabilities $\alpha_l(\omega,q,k)$, 
$\beta_l^{(0)}(\omega,q,k)$ and $\beta_l^{(1)}(\omega,q,k)$, respectively.}
\label{fig.angular1}
\end{figure} 
The arguments presented above are illustrated  by 
figure \ref{fig.angular1} where the profiles of
$\d^2 \sigma_{\pol}(\om,\Om)$  
calculated for a 3 GeV proton collision with low-$Z$
(Al$^{+12}$) and high-$Z$ (Au$^{+78}$) hydrogen-like ions.
For each ion the calculations  were performed 
for two photon energies equal to $1.5I$ and $4I$ ($I$ stands for the 
ionization potential of the target).
In these plots the length of the segment connecting
the origin and a curve point equals the value of 
$\d^2 \sigma_{\pol}(\om,\Om)$
(in millibarn/srad) in the corresponding direction.
The horizontal axis ($\theta=0$)  
is directed along the initial momentum $\bfp_1$.
The thick solid curve stand for a full relativistic 
calculation of the cross section (\ref{RBA.117}).
The contributions to $\d^2 \sigma_{\pol}(\om,\Om)$ 
from the terms proportional containing
$\left|\beta_l^{(0)}(\omega,q,k)\right|^2$,
$\left|\beta_l^{(1)}(\omega,q,k)\right|^2$,
 and 
$\left|\alpha_l(\omega,q,k)\right|^2$ is plotted
are also shown in the figure.
Note, that the sum of these contribution is not equal
to $\d^2 \sigma_{\pol}(\om,\Om)$ which, in addition, contains
the contribution of the cross-terms.
The thin solid curve in each graphs 
represents the non-relativistic dipole-photon cross section 
(\ref{RBA.117a}).

It is seen that 
in contrast to the symmetric shape of $\d^2 \sigma_{\pol}^{\NR}(\om,\Om)$,
the relativistic angular distribution is asymmetric,
being enhanced in the forward direction. 
The asymmetry increases with  $\om$ and $Z$ .

More detailed analysis of relativistic and non-dipole effects in the 
spectral and spectral-angular distributions is carried out 
in  \cite{OurRelativisticJETP}.

\section{BrS in collision of two relativistic 
complex particles.\label{AtomAtom}}

In this section we outline the specific features
which appear in the BrS process when both of the colliders, 
a projectile and a target, are complex particles (atoms, ions).
It is assumed that the relative motion of the colliders occurs
with relativistic velocities (in contrast to section 4 in 
\cite{NonRelBrS_2004} where the non-relativistic collisions were
considered), whereas the internal dynamics of the colliders 
is described in non-relativistic terms.
The theory of this process was developed in
\cite{AmusiaKuchievSolovyov1987,AmusiaSolovyov1988,AmusiaSolovyov1990a}
where one can find the details of the formalism as well as the analysis
of a number of physical phenomena and the limiting cases.

In what follows we mark all the quantities referring to a projectile
atom (or ion) with the index `1', the index  `2' is designated for 
a target atom/ion which is at rest (in the laboratory frame) before the
collision.

The atomic system of units is used.

\begin{figure}
\begin{center}
\includegraphics[scale=0.7]{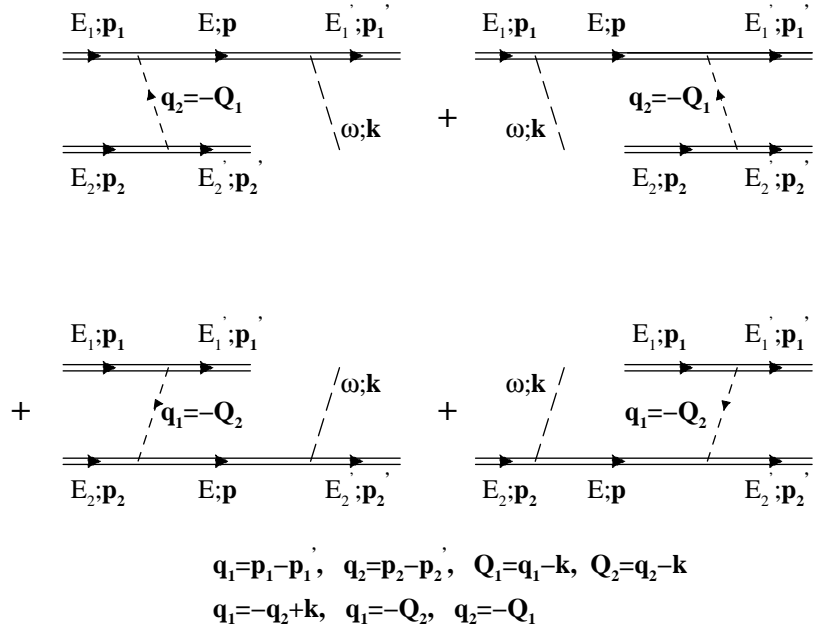}
\end{center}
\caption{Diagrammatical representation of the BrS amplitude 
in relativistic atom-atom collision.}
\label{fig:diag_at_rel}
\end{figure}
The Feynman diagrams describing the amplitude of the process are 
presented in figure \ref{fig:diag_at_rel}.
The upper double line in each diagram refers to the projectile,
the lower line describes the target, the dashed line marked with
$(\om,\bfk)$ stands for the emitted photon, and the inner dashed
line describes the (retarded) interaction between the colliders.
The initial, intermediate and final states of each collider 
are characterized with the energy (this includes the rest energy, the kinetic
energy and the energy due to the internal dynamics)
and the momentum.
The two upper diagrams describe the amplitude $f_1$ of the
photon emission by a projectile which is virtually polarized by the target,
the lower pair of diagrams represents $f_2$, which is the 
amplitude of the  photon emission by the target.
The total BrS amplitude is given by the sum $f_1+f_2$.

In contract to the non-relativistic case (see section 4 in
\cite{NonRelBrS_2004} or the original papers 
\cite{AmusiaKuchievSolovyov1984,AmusiaKuchievSolovyov1985}),
where the analytic expression for $f_1$ can be obtained from that for 
$f_2$ by exchanging the characteristics of the atoms `1' and `2',
in the relativistic collisions one should take care on the Lorenz
transformations of various quantities which constitute the amplitudes.
Therefore, to construct the amplitudes it is convenient, from the very
beginning, to operate with the four-vectors and four-tensors which allow
one to carry out all necessary transformation of the quantities using the
standard rules of quantum electrodynamics \cite{Land4}.
Then, the amplitudes $f_1$  and $f_2$ can be written in the covariant
form:
\begin{equation}
\cases{
f_2 = J_{\mu}^{(1)}(\bfq_1)\, D^{\mu\nu}(E_1-E_1^{\prime},\bfq_1)\,
T_{\lambda\nu}^{(2)}(\om, \bfk,\bfQ_2)\,e^{\lambda}
&\\
f_1 = J_{\mu}^{(2)}(\bfq_2)\, D^{\mu\nu}(E_2-E_2^{\prime},\bfq_2)\,
T_{\lambda\nu}^{(1)}(\om, \bfk,\bfQ_1)\,e^{\lambda}\, .
}
\label{v94p74n3}
\end{equation}
Here $e^{\lambda}$ is the 4-vector of the  photon polarization,
$J_{\mu}^{(j)}(\bfq_j)$ ($j=1,2$) is the four-current of the $j$th
collider due to the interaction with the virtual emission, 
$T_{\lambda\nu}^{(j)}(\om, \bfk,\bfQ_j)$ is the four-tensor
which describes the dynamic reaction of the particle to the 
action of the fields of the emitted and the virtual photons,
and  $D^{\mu\nu}(E_j-E_j^{\prime},\bfq_j)$ is the photon propagator.
The momenta $\bfq_j$ and $\bfQ_j$ are explained in 
figure \ref{fig:diag_at_rel}.

To calculate the quantities $J_{\mu}^{(j)}(\bfq_j)$ and 
$T_{\lambda\nu}^{(j)}(\om, \bfk,\bfQ_j)$ one can first consider
them in the rest frame of the $j$th collider, and then recalculate
into the laboratory frame using the Lorenz transformations rules.
This formalism is described in detail in 
\cite{AmusiaKuchievSolovyov1987,AmusiaSolovyov1988,AmusiaSolovyov1990a},
where a careful analysis of the kinematic domains, which are of the most
importance for the BrS process, was carried out as well.
Omitting the details of the evaluation and the analysis we 
present the final formulae for $f_{1,2}$.
These look differently in the cases: 
(a) a collision of two neutral complexes (e.g., atom-atom),
and
(b) a collision involving an ionic complex (e.g., ion-atom or 
ion-ion). 
The physical nature of the differences in the formula will be given
below.

In atom-atom collision the BrS amplitudes read
\begin{eqnarray}
\cases{
f_2
=
{4\pi \over c}\,
{Z_1-F^{(1)}(q_1^{\perp}) \over (q_1^{\perp})^2}\,
(\bfe\bfq_1^{\perp})\, 
\om\,\alpha^{(2)}(\om, q_1^{\perp})
&\\
f_1
=
{4\pi \over c}\,
{Z_2-F^{(2)}(q_1^{\perp}) \over (q_1^{\perp})^2}\,
\Bigl[
\tom\,\bfe\bfq_1^{\perp}
+
\gamma (\bfe\bfv_1)(\bfk\bfq_1^{\perp})
\Bigr]
\alpha^{(1)}(\tom,q_1^{\perp}),
}
\label{v94p74n15}
\end{eqnarray}
The ion-atom/ion collision is described by
\begin{eqnarray}
\cases{
f_2
=
{4\pi \over c}\,
{\cZ_1 \over q_1^2- \om^2/ c^2}
\left[
\bfe\bfq_1- {\om \over c^2}\, \bfe\bfv_1
\right]
\om\, \alpha^{(2)}_{\d}(\om),
\\
f_1 
=
{4\pi \over c}\,
{\cZ_2 \over q_2^2}
\left[
\tom\, \bfe\bfq_2^{\perp}
+
\gamma (\bfe\bfv_1)(\bfk\bfq_2^{\perp})
-
{\tom\over \gamma^2}\, \bfe\bfq_1^{\parallel}
\right]
\alpha^{(1)}_{\d}(\tom)\,.
}
\label{v94p74n17}
\end{eqnarray}
In these formulae $\bfq_j^{\perp}$ and $\bfq_j^{\parallel}$
stand for the perpendicular (`$\perp$') and the parallel 
(`$\parallel$') components of $\bfq_j$ with respect to the 
initial velocity $\bfv_1$ of the projectile,
$\gamma= (1-v_1^2/c^2)^{1/2}$ is the relativistic Lorenz factor.
The quantity $\om$ is the photon energy measured in the laboratory
frame (the rest frame of the target).
The photon emitted by the projectile and measured in its rest frame
also has energy $\om$, however, due to the Doppler effect, in the laboratory
frame this energy is changed according to the rule
$\om \to \tom=\gamma \om(1-\beta\cos\theta)$, where
$\beta = v_1/c$ and $\theta$ is the emission angle in the 
laboratory frame.
The notations $Z_j$ stand for the charges of the nucleus,
$\cZ_j$ (in (\ref{v94p74n17})) are the net ionic charges,
$F^{(j)}(q)$ denote the form-factors.
The functions $\alpha^{(j)}(\om,q)$ are non-relativistic 
 generalized dynamic dipole polarizabilities of the colliders,
$\alpha^{(1)}_{\d}(\om)$ are dynamic dipole polarizabilities.

The amplitude $f_2$ from (\ref{v94p74n15}) coincides with its 
non-relativistic analogue (see \cite{NonRelBrS_2004}, first term in 
Eq. (31)).
This coincidence has a clear physical explanation. 
The BrS process in the collision of two neutral atoms 
occurs, mainly, when the distance between the two nuclei
is less than the radius of each atom, $r \leq R_{\rm at}$.
At such distances the effect of retardation is small.
Additionally, it can be shown 
\cite{AmusiaKuchievSolovyov1987,AmusiaSolovyov1988}
that in such collisions the parallel component of the transferred
momentum is small, so that $\bfq\approx \bfq_1^{\perp}$.

The component $\bfq_1^{\perp}$ is not affected by the Lorenz
transformation with respect to the velocity $\bfv_1$.
Therefore, as it is seen from (\ref{v94p74n15}), the amplitude 
$f_1$ can be obtained from $f_2$ by choosing the frame moving with the
velocity $\bfv_1$ (instead of the laboratory frame) and, then, exchanging
the colliders indices. 

The BrS process in ion-ion (or ion-atom) collisions is 
governed by another regime \cite{AmusiaSolovyov1990a}.
In this case, due to the presence of the non-zero net charges,
the colliders can be effectively polarized even
being separated by a large distance, $r \gg R_{\rm at}$.
Therefore, the retardation becomes very important.
 This explains the factor $q_1^2- \om^2/ c^2$ in the denominator
in the first expression from (\ref{v94p74n17}).
The dominating role of large distances is also reflected 
by the fact that the amplitudes $f_1$ and $f_2$ are expressed in terms
of the polarizabilities $\alpha^{(1)}_{\d}(\tom)$ and 
$\alpha^{(2)}_{\d}(\om)$ rather than via the  generalized polarizabilities
as in (\ref{v94p74n15}).
Additionally, in contrast to the neutral atoms collision,
both components, the transverse and the parallel, of the vectors
$\bfq_j$ are important in ionic collisions, which explains the
additional terms in square brackets in (\ref{v94p74n17}).

Formulae (\ref{v94p74n15}) and (\ref{v94p74n17}) allow one
to analyze the spectral and spectral-angular dependence
of BrS formed in collision of relativistic complex particles
\cite{AmusiaKuchievSolovyov1987,AmusiaSolovyov1988,AmusiaSolovyov1990a}.
These dependences include the contributions of the radiation 
by the target, the projectile and the interference term.
Not going to reproduce rather long expressions we mention
the most important features of these distributions.

For all types of the collision the leading terms in the 
angular distribution of the dipole PBrS emitted by the target 
and the projectile can be presented in the 
following general forms:
\begin{eqnarray}
\cases{
{\d^2\sigma_2 \over \d\om \d\Om} 
=
C_2\, (1+\cos^2\theta)
\\
{\d^2\sigma_1 \over \d\om \d\Om}
=
C_1\,(1+\cos^2 \tth) 
\left({\tom \over \om}\right)^2\, ,
}
\label{v94p74n22}
\end{eqnarray}
where the coefficients $C_{1,2}$ are independent on 
the emission angle $\theta$.
Omitting details we mention that for atom-atom
collisions these coefficients weakly depend on the projectile energy, whereas
in the ion-ion/atom case they logarithmically increase with 
$\gamma=E/M_1c^2$.
This happens due to the same reasons which are mentioned above
in connection with the PBrS of a structureless particle 
(see (\ref{RBA.45}), (\ref{RBA.45c}) and figure \ref{fig_E-dep}).

It is seen from (\ref{v94p74n22}), 
that the radiation by the target, $\d^2\sigma_2/\d\om \d\Om$,
is proportional to $(1+\cos^2\theta)$, i.e. is distributed as 
that emitted by a rotating dipole \cite{Landau2}.
This is a common feature and it does not depend either on 
the type of the projectile (light, heavy, structureless or complex) nor
on the type of the interaction between the colliders.
Apart from the collisions of the atomic particles 
this dipole-type profile of the angular distribution was found in 
a neutron and  a neutrino BrS process in collisions with atoms
\cite{AmusiaBaltenkovZhalovKorolSolovyov1986,AmusiaBaltenkovKorolSolovyov1987},
and in nuclear collisions \cite{HubbardRose1966,AmusiaSolovyovNuclear1987}.
Beyond the non-relativistic dipole approximation and in the case when
the internal dynamics of the target must be treated relativistically,
the profile of the angular distribution strongly deviates from the 
$(1+\cos^2\theta)$ law (see figures \ref{fig.angular1}c,d).

The PBrS of the projectile, $\d^2\sigma_1/\d\om \d\Om$, has another
angular dependence.
As mentioned above, the photon energy  emitted by the projectile is 
Doppler-shifted, $\om \to \tom$. 
This results in the factor $(\tom/\om)^2\propto (1-\beta\cos\theta)^2$
which strongly deviates from one in the case of high velocities, when
$\beta=v_1/c\sim 1$.
Additional modification of the angular dependence is due to the 
abberation effect.
The quantity $\tth$, on the right-hand side of the second equation from
(\ref{v94p74n22}), stands for the emission angle in the rest frame of the
projectile.
In the laboratory frame this angle is becomes abberated and is related to
$\theta$ via 
$\cos\tth = (\cos\theta-\beta)/(1- \beta \cos \theta)$
 \cite{Landau2}.
Therefore, the profile of $\d^2\sigma_1/\d\om \d\Om$ is defined by 
$\cP(\theta,\beta)\equiv (1+\beta^2)(1+\cos^2 \theta) - 4\beta\cos\theta$.
This function explicitly demonstrates a remarkable feature of the 
PBrS by a relativistic complex projectile.
Namely, it does not contain the well-known peculiarity typical for the
ordinary BrS, where the radiation is concentrated in a narrow cone 
$\theta \leq \gamma^{-1}$ in the forward direction.
Indeed, the function $\cP(\theta,\beta)$ exhibit an opposite property:
the radiation emitted in the forward direction, $\theta=0$, 
is less intensive than in the backward direction, $\theta=\pi$.
In the non-relativistic limit the function $\cP(\theta,\beta)$ reduces
to the dipole-type  profile $\cP(\theta,0)=(1+\cos^2\theta)$ 
\cite{AmusiaKuchievSolovyov1985}.

On the basis of the detailed analysis carried in
\cite{AmusiaSolovyov1988,AmusiaSolovyov1990a} the following criterion
was formulated on the magnitude of the ratio 
$\xi = (\d^2\sigma_1/\d\om \d\Om)/(\d^2\sigma_1/\d\om \d\Om)$ 
in different ranges of $\om$ and $\theta$.
It was shown, that if the photon energy and the emission angle
are chosen to satisfy the inequality
$(\tom/c)^2\left|\alpha^{(1)}(\tom, q_1^\perp)\right|^2 \gg
\left|\alpha^{(2)}(\om, q_1^\perp)\right|^2$,
then the PBrS of the projectile dominates in the total spectrum over 
the radiation emitted by the target.
In the inequality sign is opposite then the target radiates more
intensively.
Therefore, for sufficiently high velocities of the collision there is
a possibility to separate the radiation by the two colliders.

Another important feature of the BrS process at relativistic velocities,
which is due to the Doppler and the abberation effects, is
that the intensity of the dipole radiation formed in symmetric collisions
does not vanish.
Indeed, if one considers two identical colliders then
the total amplitude of the BrS $f_1+f_2 \neq 0$ (see (\ref{v94p74n17})).
We note that in the non-relativistic symmetric collision the emission
of the dipole photon is strictly forbidden and $f_1+f_2 = 0$
(see section 4 in \cite{NonRelBrS_2004} or the original papers 
\cite{AmusiaKuchievSolovyov1984,AmusiaKuchievSolovyov1985}).

\section{Conclusions.\label{Conclusions}}

In this paper reviewed the progress which have been achieved
in theoretical description of the PBrS process with account for 
the relativistic effects.
We concentrated mainly on the polarizational part of the total
BrS spectrum.
This mechanism defines the emission spectrum formed in 
collisions of heavy projectiles with many-electron targets.
In the case of a sctructureless heavy projectile 
the main characteristics of the spectrum can be accurately
computed using the algorithms which have been developed 
recently on the basis of the formalism described in section
\ref{Structureless}.
The main difficulty, on the numerical level,
is in the accurate and efficient calculation of the 
relativistic generalized polarizabilities of a many-electron target.
Compared to  the non-relativistic case this is a more 
complicated problem, and at present we cannot state that these 
quantities can be efficiently computed in arbitrary ranges of
the photon energies, transferred momentum and the multipolarity index.
 However, the calculations which have been already performed 
indicated that the relativistic effects of all types, --
i.e. those related to the motion of the colliders, including 
the internal dynamics, the effect of retardation and the
radiation in multipoles, 
must be accounted for in order to obtain reliable results.

Note that if a relativistic projectile has the internal structure
of a relativistic nature, it is not necessary to develop a new 
formalism describing the PBrS emitted by the projectile.
Instead, the corresponding formulae can be obtained
from those presented \ref{Structureless} using the Doppler and 
the aberration of light transformation as it is described in
\ref{AtomAtom}.
With slight  modifications the formalism discussed  above
can be applied for other colliding systems, where
relativistic effects are important. 
Calculations for relativistic  heavy-ion collisions are of interest
because of recent experimental efforts in this direction 
\cite{LudziejewskiEtal1998}.
For example, one can  describe the BrS arising in relativistic 
collisions involving nuclei.
In this case the dynamic polarization of the colliders results in 
the photon emission via the PBrS mechanism, and the main contribution 
comes from the non-dipole radiation (quadrupole and higher).

For a light projectile, an electron or a positron, 
the total BrS problem cannot be considered in terms of 
the polarizationa mechanism only.
The full theory must include, from the very beginning, the two
terms in the amplitude, $f_{\ord}$ and $f_{\tot}$.
As a result, the total BrS cross section $\d\sigma_{\tot}$
includes the ordinary $\d\sigma_{\ord}$ 
and the polarizational $\d\sigma_{\pol}$ parts, which are positive,
and the interference term $\d\sigma_{\rint}$ which can be of either
sign.
In the non-relativistic domain and within the dipole-photon scheme
the features of $\d\sigma_{\tot}$ are known quite well
(see, e.g. (\cite{NonRelBrS_2004})).
From the formal viewpoint the theory of the BrS process of a 
light projectile, which incorporates all relativistic effects,
has been developed.
Indeed, the formalism for the PBrS presented in 
\cite{OurRelativisticJPB,OurRelativisticJETP} and sketched above
in section  \ref{Structureless} conbined with that for the ordinary
BrS \cite{PrattTseng1970} (see also \cite{ShafferPratt1997}
where the relativistic DPWA is compared with simpler theories)
allows one, in principle, to obtain the characteristics of the
total BrS.
The calculation, within fully relativistic scheme, of the 
cross section $\d\sigma_{\pol}$ of an electron now can be
implemented. 
For example, the data presented in section  \ref{Structureless}
for a 3 GeV positron can be also attributed for an electron
of the energy $\E_1\approx 1.5$ MeV, since the PBrS part of the
spectrum is nearly independent on the mass of a projectile.
Thus, one can estimate the total cross section as a sun of the two
terms $\d\sigma_{\tot}\approx \d\sigma_{\ord}+\d\sigma_{\pol}$ where 
$\d\sigma_{\ord}$ can be taken from the tables \cite{AtData1,AtData2}.
This approach completely ignores the interference term
$\d\sigma_{\rint}$.
The calculation of the total BrS cross section based on the
approximation $\d\sigma_{\tot}\approx \d\sigma_{\ord}+\d\sigma_{\pol}$
were carried out recently in \cite{AstapenkoBureevaLisitsa2000b}.
This was done within the `logarithmic approximation' and for the 
dipole-photon radiation only. The relativistic effects due to the
internal dynamics of the taget electrons were not accounted for as well.
However, it is not at all evident that the term $\d\sigma_{\rint}$ 
can be ignored on the basis of simple approaches.
As figures \ref{fig.angular1} demonstrate, the relativistic effects
noticeably modify the angular distribution of PBrS leading to the
increase of the emission in the forward direction.
Since the OBrS of a relativistic projectile is also emitted 
mostly in the  forward direction the interference of the two
mechanisms can be important.
The approach which allows one to account for the interference 
was proposed in \cite{AvdoninaPratt1999}, where analytic expressions 
and the numerical results for the total BrS spectra from neutral 
atoms and ions were presented for electron scattering at 
energies $10-2000$ keV. 
In the cited paper the 'stripping approximation', derived initially
\cite{BuimistrovTrakhtenberg1977,AmusiaAvdoninaChernyshevaKuchiev1985}
for a non-relativistic electron--atom scattering,
was extended to the case of the relativistic velocities of the collision.
On the basis of this model approach (which does not accurately
account for the retardation effects and for the multipole character of the 
radiation) 
the analysis of the modification of the BrS spectrum due to the
influence of the PBrS channel was carried out.

At present, there are no accurate numerical results on the total BrS
spectra and angular distributions of relativistic electrons/positron 
on many-electron targets which account for all important relativistic
effects and for the two mechanisms of the photon emission.
Let us stress that such calculations will be an important step
forward towards precise comparison with the recent experimental data
on the BrS obtained for collisions of 
$10-100$ keV electrons with various targets 
\cite{QuarlesPortillo1999}.
The mostly recent experiments \cite{PortilloQuarles_PRL}, 
where for the first time the measurement of the absolute values of
the BrS cross section in electron scattering from noble gas atoms were 
reported, show some indications on the significance of the PBrS
in the energy ranges in which the relativistic treatment of the process
is necessary.

\ack

This work is supported  by the Russian Foundation for Basic Research
(Grant No 96-02-17922-a) and INTAS (Grant No 03-51-6170).
AVK acknowledges the support from the Alexander von Humboldt Foundation.

\section*{References}


\begin{thebibliography}{99}

\bibitem{NonRelBrS_2004}
        Korol, A.~V., Solov'yov, A.~V.,
         2005. 
         Radiat. Phys. Chem. this issue.
\bibitem{Akhiezer}
        Akhiezer, A.~I., Berestetskii V.~B.,
        1969.
        Quantum Electrodynamics. Nauka, Moscow.
\bibitem{Landau2}
        Landau, L.~D., Lifshitz, E.~M.,
        1975.
        The Classical Theory of Fields. Pergamon, Oxford.
\bibitem{AmusiaKorolKuchievSolovyov1985}
        Amusia, M.~Ya., Kuchiev, M.~Yu., Korol, A.~V., Solov'yov, A.~V.,
        1985. Sov. Phys. - JETP 61, 224-228.
\bibitem{AstapenkoBuimistrovKrotovMikhailovTrahtenberg1985}
        Astapenko, V. A., Buimistrov, V. M., Krotov, Yu. A.,
        Mihailov, L.~K., Trakhtenberg, L.~I.,
        1985.
         Sov. Phys. -- JETP 61,   930.
\bibitem{AmusiaKuchievSolovyov1987}
        Amusia, M.~Ya., Kuchiev, M.~Yu., Solov'yov, A.~V.,
        1987. Sov. Phys. - Tech. Phys. 32, 499-500.
\bibitem{AmusiaSolovyov1988}
        Amusia, M.~Ya., Kuchiev, M.~Yu., Solov'yov, A.~V.,
        1988. Sov. Phys. -- JETP  67, p.41-48.
\bibitem{AmusiaSolovyov1990a}
        Amusia, M.~Ya., Solov'yov, A.~V.,
        1990. Sov. Phys. -- JETP 70, 416-425.
\bibitem{AvdoninaPratt1999}
         Avdonina, N.~B., Pratt, R.~H.,
        1999.
        J.~Phys.~B 32, 4261-4276.
\bibitem{AstapenkoBureevaLisitsa2000b}
        Astapenko, V.~A., Bureeva, L.~A., Lisitsa, V.~S.,
        2000.
        JETP 90,  788-794.
\bibitem{KorolLyalinObolenskySolovyovSolovjev2001}
         Korol, A.~V., Lyalin,A.~G., Obolenski, O.~I., Solov'yov, A.~V., 
         Solovjev, I.A., 
         2001.
         In: Dugganm J.~L., Morgan, I.~L. (Eds.)
         AIP Conference Proceedings, vol. 576.
         AIP Press,  pp. 64-67.
\bibitem{OurRelativisticJPB}
        Korol, A.~V., Obolensky, O.~I., Solov'yov, A.~V., Solovjev, I.~A.,
        2001.
        J. Phys. B 34, 1589-1617.
\bibitem{OurRelativisticJETP}
        Korol, A.~V., Lyalin, A.~G., Obolensky, O.~I., 
        Solov'yov, A.~V., Solovjev, I.~A.,
        2002.
        JETP  94, 704-719.
\bibitem{KorolObolenskySolovyovSolovjev2002}
        Korol, A.~V., Obolensky, O.~I., Solov'yov, A.~V., Solovjev, I.~A.,
        2002. Surface Review and Letters  9, 1191-1195.
\bibitem{BlazhevichEtal1996}
        Blazhevich, S.V.,  Cherpunov, A.S., Grishin, V.K., et. al.,
        1996. Phys.~Lett. A 211, 309-312.
\bibitem{Nasonov1998}
        Nasonov, N.~N.,
        1998.
        Nucl. Instrum. Methods B 1998 145, 19-24.
\bibitem{BlazhevichEtal1999}
        Blazhevich, S.V.,  Cherpunov, A.S., Grishin, V.K., et. al.,
        1999. Phys.~Lett. A 254, 230-232.
\bibitem{KamyshanchenkoNasonovPokhil2001}
        Kamyshanchenko, N., Nasonov, N., Pokhil, G.,
        2001.
        Nucl. Instrum. Methods B  173, 195-202.
\bibitem{AstapenkoBuimistrovKrotovNasonov2004}
        Astapenko, V.~A., Buimistrov, V.~M., Krotov, Yu.~A., 
        Nasonov, N.~N.,
        2004.
        Phys. Lett. A 332,  298-302.
\bibitem{AstapenkoBureevaLisitsa2002}
         Astapenko, V.~A., Bureeva, L.~A., Lisitsa, V.~S.,
         2002.
         Phys. Usp. 45, 149-184.
\bibitem{Zon1977}
         Zon, B.~A.,
         1977.
         Sov. Phys. -- JETP 46, 65.
\bibitem{AmusiaZimkinaKuchiev1982}
         Amusia M.~Ya., Zimkina T.~M., Kuchiev M.~Yu.,
        1982.
        Sov. Phys. - Tech. Phys. 27, 866.
\bibitem{AmusiaAvdoninaChernyshevaKuchiev1985} 
        Amusia, M.~Ya., Avdonina, N.~B., Chernysheva, L.~V., Kuchiev, M.~Yu.,
        1985.
        J. Phys. B 18, L791-L796.
\bibitem{BuimistrovTrakhtenberg1977}
         Buimistrov, V.M., Trakhtenberg, L.~I.,
         1977.
         Sov. Phys. -- JETP 46, 447.
\bibitem{Land4}
       Berestetskii, V.~B., Lifshitz, E.~M., Pitaevskii, L.P.,
       1982.
       Quantum Electrodynamics. Pergamon, Oxford.
\bibitem{PrattTseng1970}
         Tseng, H. K., Pratt, R. H.,
         1970.
         Phys. Rev. A 3 100-115.
\bibitem{Tseng1997}
         Tseng, H.~K.,
         1997.
         J. Phys. B 30,  L317-L321.
         (Corrigendum:  2000. ibid. 33, 1471.)
\bibitem{Korol1992}
         Korol, A.~V., 
         1992. J. Phys. B  25, L341-L344.
\bibitem{KorolLyalinSolovyovAvdoninaPratt2002}
        Korol, A.~V., Lyalin, A.~G., Solovyov, A.~V., Avdonina, N.~B.,
        Pratt, R.~H.,
        2002.
        J. Phys. B 35, 1197-1210.
\bibitem{VarshalovichMoskalevKhersonskii}
         Varshalovich, D.~A., Moskalev, A.~N., Khersonskii, V.~K.,
        1988.
        Quantum Theory of Angular Momentum.
        World Scientific, Singapore.
\bibitem{LindgrenMorrison}
         Lindgren, I., Morrison, J.,
        1986.
        Atomic Many-Body Theory.
        Springer, Berlin.
\bibitem{ChernyshevaYakhontov1999}
         Chernysheva, L.~ V., Yakhontov, V.~L.,
         1999.
         Comp. Phys. Comm. 119, 232-249.

\bibitem{AmusiaKorol1992}
        Amusia, M.~Ya., Korol A.~V.,
        1992.
        J. Phys. B 25, 2383-2392.
\bibitem{AmusiaAvdoninaKuchievChernysheva1986}
        Amusia, M.~Ya., Avdonina, N.~B., Kuchiev, M.~Yu., 
        Chernysheva, L.~V., 
        1986.
        Izv. Acad. Nauk SSSR: Ser. Fiz. 50, 1261-1266.



\bibitem{AmusiaKuchievSolovyov1984}
        Amusia, M.~Ya., Kuchiev, M.~Yu., Solov'yov, A.~V.,
        1984.
        Sov. Phys. - Tech. Phys. Lett. 10, 431-432.
\bibitem{AmusiaKuchievSolovyov1985}
        Amusia, M.~Ya., Kuchiev, M.~Yu., Solov'yov, A.~V.,
        1985. Sov. Phys. -- JETP 62, 876-881.
\bibitem{AmusiaBaltenkovZhalovKorolSolovyov1986}
        Amusia, M.~Ya., Baltenkov, A.~S., Zhalov, M.~B., 
        Korol, A.~V., Solov'yov, A.~V.,
        1986.
        Polarisational bremsstrahlung in scattering of particles at atoms 
        and nuclei.- Proceedings of 21st Winter School of Leningrad 
        Institute of Nuclear Physics,  pp.135-194 (in Russian). 
\bibitem{AmusiaBaltenkovKorolSolovyov1987}
        Amusia, M.~Ya., Baltenkov, A.~S., Korol, A.~V., Solov'yov, A.~V.,
        1987. Sov. Phys. -- JETP 66, 877-883.
\bibitem{HubbardRose1966}
        Hubbard, D.~F., Rose, M.~E.,
        1966. Nucl. Phys. 84, 337.
\bibitem{AmusiaSolovyovNuclear1987}
        Amusia, M.~Ya., Solov'yov, A.~V.,
        1987. 
        In: Sushkov, O.~P. (ed.),
        Modern Developments in Nuclear Physics,
        World Scientific, Singapore, pp. 425-437.
\bibitem{LudziejewskiEtal1998}
         Ludziejewski, T., St\"ohlker, T.,  Keller, S., et. al.,
         1998. 
         J. Phys. B 31, 2601-2609.
\bibitem{ShafferPratt1997}
         Shaffer, C.~D., Pratt, R.~H.,
         1997.
         Phys.  Rev. A 56,  3653-3658.
\bibitem{AtData1}
         Pratt, R. H., Tseng, H.~K., Lee, C.~M., Kissel, L., 
         MacCallum, C., Riley, M.,
         1977.
         At Data Nucl. Data Tables 20, 175-209
         (Erratum: 1981. ibid. 26, 477-481).
\bibitem{AtData2}
         Kissel, L., Quarles, C.~A., Pratt, R.~H.,
         1983.
         At. Data Nucl. Data Tables 28, 381.
\bibitem{QuarlesPortillo1999}
         Quarles, C.~A., Portillo, S.,
         1999.
         In: Duggan J.~L., Morgan, I.~L. (Eds.)
         AIP Conference Proceedings, vol. 576.
         AIP Press,  pp. 174-177.
\bibitem{PortilloQuarles_PRL}
         Portillo, S., C. A. Quarles, C.~A.,
         2003.
         Phys. Rev. Lett. 91, 173201.


\end{thebibliography}
\end{document}